\documentclass[fleqn,usenatbib]{mnras}

\usepackage{newtxtext,newtxmath}

\usepackage{hyperref}

\usepackage[T1]{fontenc}

\DeclareRobustCommand{\VAN}[3]{#2}
\let\VANthebibliography\thebibliography
\def\thebibliography{\DeclareRobustCommand{\VAN}[3]{##3}\VANthebibliography}

\def\mearth{{\rm\,M_\oplus}}

\usepackage{graphicx}	
\usepackage{amsmath}	

\title[Co-orbital constellations]{Constellations of co-orbital planets: horseshoe dynamics, long-term stability, transit timing variations, and potential as SETI beacons}

\author[Raymond, Veras, Clement, Izidoro, Kipping \& Meadows]{Sean N. Raymond$^1$\thanks{E-mail: rayray.sean@gmail.com}, 
Dimitri Veras$^\mathrm{2,3,4}$, 
Matthew S. Clement$^{5,6}$, 
Andre Izidoro$^{7,8}$,
\newauthor David Kipping$^9$ and 
Victoria Meadows$^{10,11}$
\\
$^\mathrm{1}$Laboratoire d'Astrophysique de Bordeaux, CNRS and Universit{\'e} de Bordeaux, All{\'e}e Geoffroy St. Hilaire, 33165 Pessac, France \\
$^2$Centre for Exoplanets and Habitability, University of Warwick, Coventry CV4 7AL, UK\\
$^3$Centre for Space Domain Awareness, University of Warwick, Coventry CV4 7AL, UK\\
$^4$Department of Physics, University of Warwick, Coventry CV4 7AL, UK\\
$^5$Earth and Planets Laboratory, Carnegie Institution for Science, 5241 Broad Branch Road, NW, Washington, DC, 20015, USA\\
$^6$Johns Hopkins APL, 11100 Johns Hopkins Rd, Laurel, MD 20723, USA\\
$^7$Department of Physics and Astronomy, 6100 Main MS-550, Rice University, Houston, TX 77005, USA \\
$^8$Department of Earth, Environmental and Planetary Sciences, MS 126, Rice University, Houston, TX 77005, USA\\
$^9$Department of Astronomy, Columbia University, 550 West 120th Street, New York, NY 10027, USA\\
$^{10}$Department of Astronomy and Astrobiology Program, University of Washington, Box 351580, Seattle, WA 98195, USA\\ 
$^{11}$NASA Astrobiology Institute's Virtual Planetary Laboratory, Box 351580, University of Washington, Seattle, WA 98195, USA
}

\date{Accepted XXX. Received YYY; in original form ZZZ}

\pubyear{2022}

\begin{document}
\label{firstpage}
\pagerange{\pageref{firstpage}--\pageref{lastpage}}
\maketitle

\begin{abstract}
Co-orbital systems contain two or more bodies sharing the same orbit around a planet or star. The best-known flavors of co-orbital systems are tadpoles (in which two bodies' angular separations oscillate about the L4/L5 Lagrange points $60^\circ$ apart) and horseshoes (with two bodies periodically exchanging orbital energy to trace out a horseshoe shape in a co-rotating frame). Here, we use $N$-body simulations to explore the parameter space of many-planet horseshoe systems. We show that up to 24 equal-mass, Earth-mass planets can share the same orbit at 1 au, following a complex pattern in which neighboring planets undergo horseshoe oscillations. We explore the dynamics of horseshoe constellations, and show that they can remain stable for billions of years and even persist through their stars' post-main sequence evolution. With sufficient observations, they can be identified through their large-amplitude, correlated transit timing variations. Given their longevity and exotic orbital architectures, horseshoe constellations may represent potential SETI beacons.

\end{abstract}




\section{Introduction}

Our Solar System contains two different types of co-orbital configurations: horseshoes and tadpoles.\footnote{Other types of co-orbital systems have been found to exist, such as the 1:1 eccentric resonance~\citep{laughlin02}, but no examples have been found in the present-day Solar System. } Jupiter's Greek and Trojan asteroids have tadpole orbits and librate about the L4 and L5 Lagrange points, 60 degrees ahead of and behind Jupiter, respectively. They contain an integrated mass of ${\sim}10^{-5} \mearth$, roughly 2\% as much mass as the main asteroid belt~\citep{vinogradova15,demeo13}.  Jupiter's tadpole co-orbitals are confined to a restricted region of phase space such that the Greeks and Trojans always remain on separate sides of Jupiter~\citep[e.g.][]{stacey08}.  Trojan companions have also been detected co-orbiting with Venus~\citep[e.g.][]{delafuente14}, Earth~\citep{wiegert97,connors11,santana-ros22}, Mars~\citep[e.g.][]{mikkola94}, Saturn~\citep{alexandersen21}, Uranus~\citep{delafuente15} and Neptune~\citep{sheppard06}.\footnote{The Minor Planet Center maintains a list of Trojans of the planets at \url{https://www.minorplanetcenter.net/iau/lists/Trojans.html}.}

Satellites Janus and Epimetheus orbit Saturn in a horseshoe co-orbital configuration~\citep{smith80,dermott81a,dermott81b}. The two moons orbit at very slightly different orbital radii, with slightly different mean motion, such that they undergo periodic close encounters every four years.  During each encounter, Janus and Epimetheus exchange orbital energy and swap positions relative to Saturn. Being less massive and carrying less orbital energy, Epimetheus experiences a larger shift in orbital radius than Janus. Epimetheus therefore has a larger angular velocity relative to a frame co-moving at the moons' mean orbital radius, and when viewed in that frame it traces out a horseshoe-shape. Meanwhile, Janus undergoes a smaller shift and undergoes a much lower-amplitude oscillation between encounters with Epimetheus. 


The Solar System's co-orbital configurations include small bodies (Trojan asteroids) and planetary satellites (Janus and Epimetheus), but there are no co-orbital planets.  Nonetheless, it is thought that systems of co-orbital systems with two planets should exist.  Simulations of planet formation and migration have shown that Trojan planets -- two planets orbiting in or near their common L4/L5 points -- should form readily and survive~\citep[e.g.][]{beauge07,cresswell09,rodriguez19}. Cohorts of Earth- to Neptune-mass planets migrating inward together often scatter each other onto crossing orbits and, through the damping action of the gas disk, are trapped as Trojans~\citep{cresswell09,izidoro17,raymond18b}. Migration can also produce horseshoe configurations~\citep{rodriguez19}.  Dynamical instabilities among giant planets can also occasionally produce co-orbital planets; for instance one simulation from \cite{clement21} created an ice giant Trojan pair during the Solar System's giant planet instability.  Earth-mass planets can directly accrete at the Lagrange points of a gas giant to produce Trojans or horseshoes~\citep{laughlin02,beauge07,lyra09,izidoro10}.  Trojan planets generally survive long-range orbital migration, except for planet masses close to the gap-opening mass threshold~\citep[which depends on the properties of the gaseous disk but is typically close to Saturn-mass;][]{pierens14b}.

Co-orbital planets should be detectable.  Several studies have shown that co-orbital exoplanet systems produce radial velocity~\citep{laughlin02,madhu09,giuppone12,leleu15} and transit timing variation~\citep{ford06b,ford07,haghighipour13,vok14,dobrovolskis15,veras16b} signals that should be detectable, although they may be hard to interpret.  While no clear signal has been detected to date~\citep[e.g.][]{rowe06} there exist ongoing searches for Trojan exoplanet systems~\citep[e.g., the TROY project:][]{lillobox18a,lillobox18b}.

In this paper, we numerically demonstrate the existence of horseshoe `constellations', in which many planets share the same orbit around a star.  In these constellations, planets continuously undergo horseshoe libration with their neighbors. We will show that horseshoe systems can have as many as 24 planets sharing a single orbit while remaining stable for billions of years, and can survive post-main sequence evolution of their central star; in our simulations we use a Sun-like central star, but the dynamics can be applied to any stellar mass.

We use a suite of N-body simulations to explore the survival and dynamics of horseshoe systems with up to 24 planets, testing different system parameters as well as their post-main sequence survival (Section 2).  We also consider the detectability of such systems via transit-timing variations (Section 3).  Finally, we discuss the formation of such systems, and their astrophysical implications as potential SETI beacons (Section 4).  


\section{Dynamics and stability of horseshoe constellations}

\subsection{N-body Simulations}
Our simulations started with two or more planets sharing the same orbit, in most cases $1 \mearth$ planets sharing a circular orbit at 1 au around a Solar-mass star.  Systems were initially spaced in units of the mutual Hill radius, $R_{H,m}$, defined as:
\begin{equation}
    R_{H,m} = a \left(\frac{m_1 + m_2}{3 M_\star}\right)^{1/3},
\end{equation}
\noindent where $a$ is the planets' semimajor axis, $m_1$ and $m_2$ are the planets' masses (with $m_1 = m_2 = 1 \mearth$ in most of our simulations), and $M_\star$ is the stellar mass of $1 M_\odot$.

We initialized our simulations with $N_p$ planets equally spaced by $\Delta$ mutual Hill radii along the same orbit.  We ran simulations with $N_p =$~2, 3, 4, 5, 6, 9, 12, 20, 22, and 24. We treated the initial inter-planet spacing as a free parameter and we tested for different values of $\Delta$ to find the spacing needed for long-term dynamical stability, for each value of $N_p$.  The chosen orbit was near-circular, with a semimajor axis of 1 au and an eccentricity of $10^{-5}$.  Given that all planets shared the same orbit, they had no mutual inclination~\citep[note that horseshoe orbits can have modest mutual inclinations and remain stable; e.g. see][]{cuk12b}.  We also ran simulations testing the effect of the planet mass and the stability of systems with non-equal mass planets (see Section 2.4). 

We integrated our systems with the hybrid integrator within the {\tt Mercury} integration package~\citep{chambers99}. We used a `hybrid integrator changeover' parameter of 3 Hill radii; this parameter determines at what distance between planets the integrator switches from the mixed-variable symplectic (MVS) map~\citep{wisdom91} to a Bulirsch-Stoer scheme.  Given the stability limit of 5 Hill radii for horseshoe orbits~\citep{cuk12b}, this means that our stable simulations were integrated with the MVS method. To verify that the integration scheme was adequate, we performed a series of tests on 3-planet systems with a Bulirsch-Stoer integrator and found very good agreement (see Appendix A).  We integrated our systems until either they became unstable, as indicated by a close encounter during which two planets passed within one Hill radius of each other, or until they reached 1 Gyr (for 2- and 3-planet systems) or 100 Myr (for larger-$N_p$ systems).

\begin{figure}
	\includegraphics[width=\columnwidth]{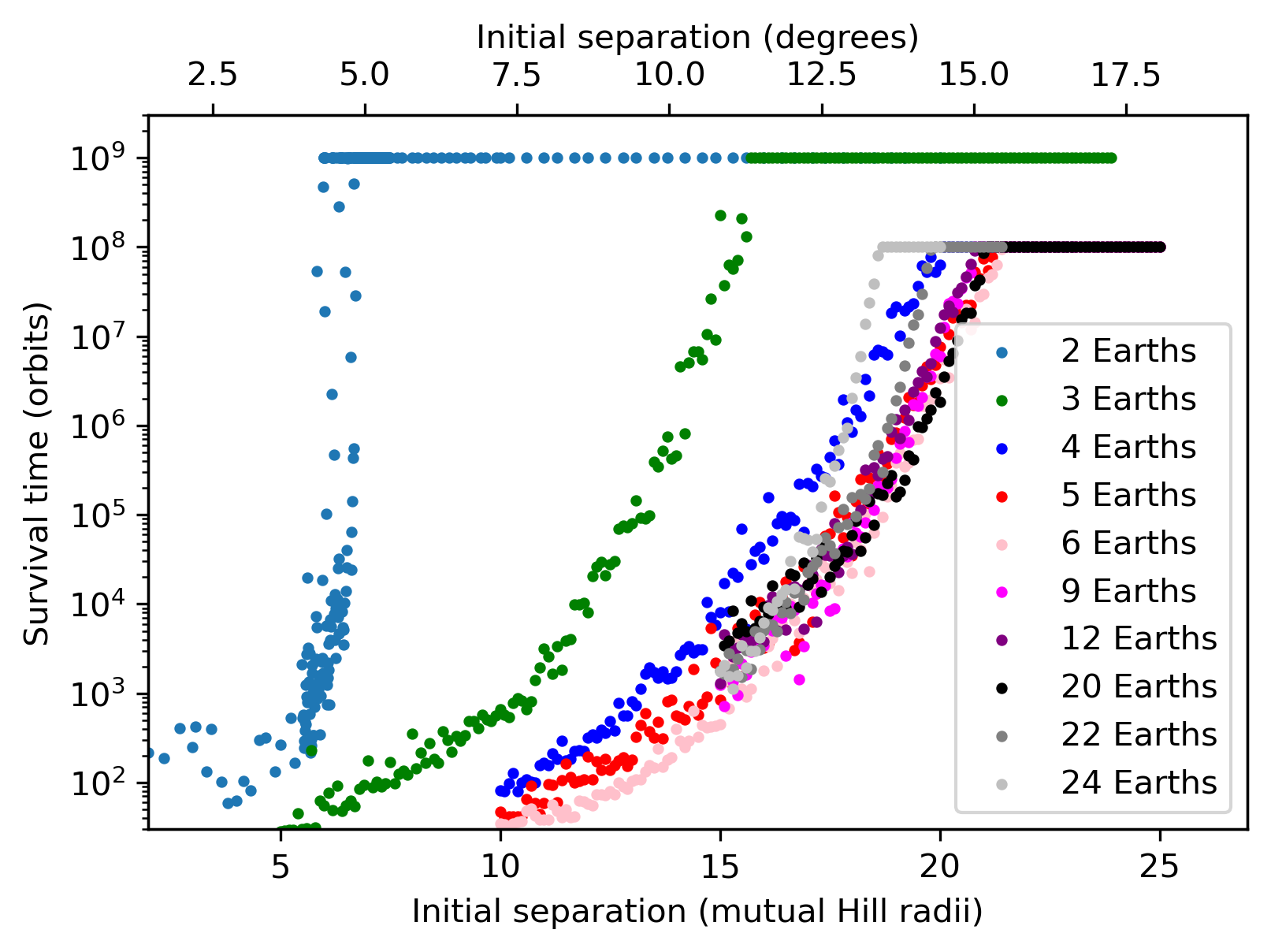}
    \caption{Survival of co-orbital systems with different numbers of Earth-mass planets as a function of the initial inter-planet separation. The plateaus at 1 Gyr (for $N_p=$~2-3 planets) and 100 Myr (for $N_p > 3$ simply represent the length of the simulations; those systems would almost certainly remain stable for far longer.  The stability of systems with 5 or more planets is roughly independent of the number of planets. }
    \label{fig:survtime}
\end{figure}

\begin{figure*}
 	\includegraphics[width=0.66\columnwidth]{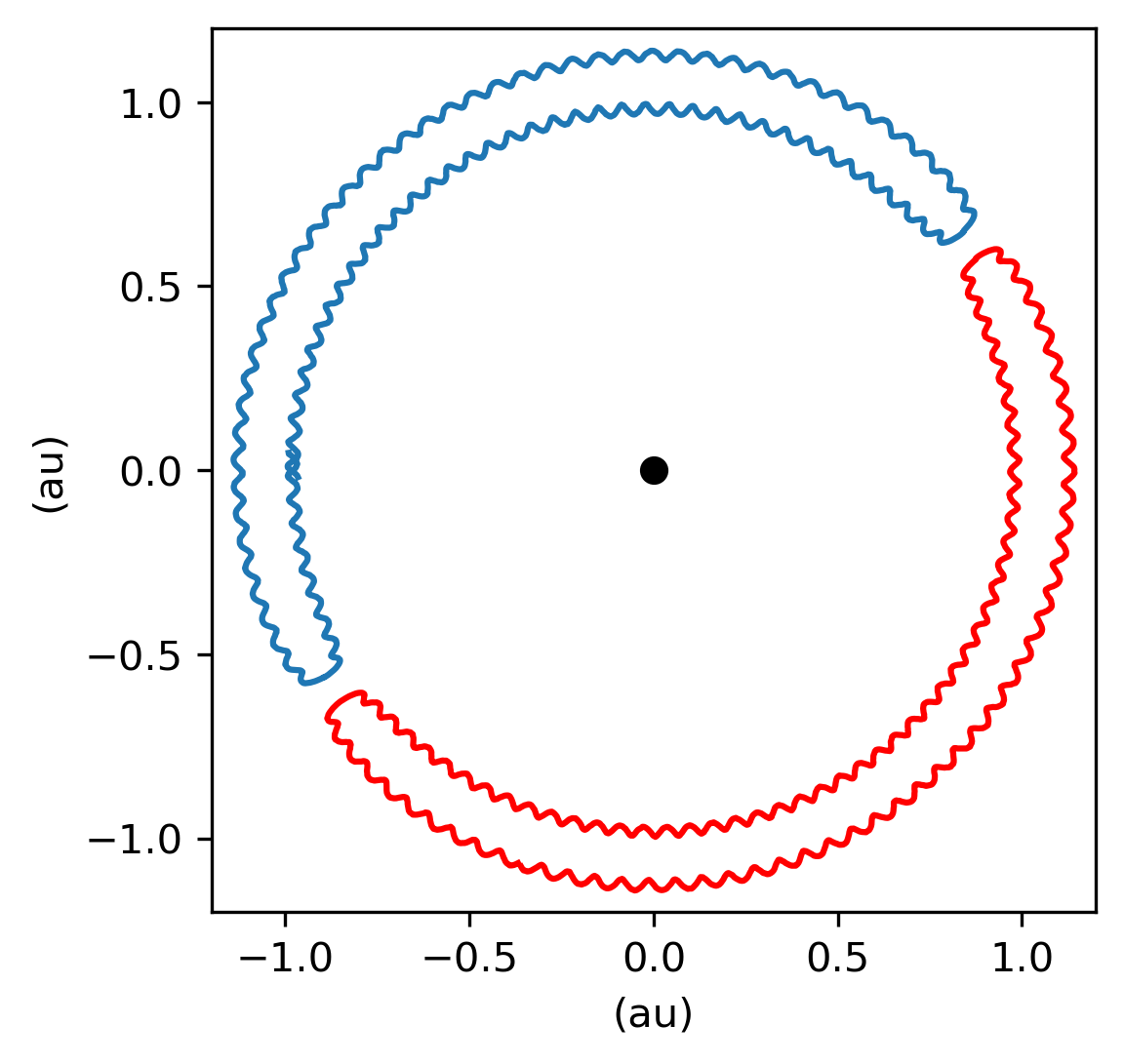} 
  	\includegraphics[width=0.66\columnwidth]{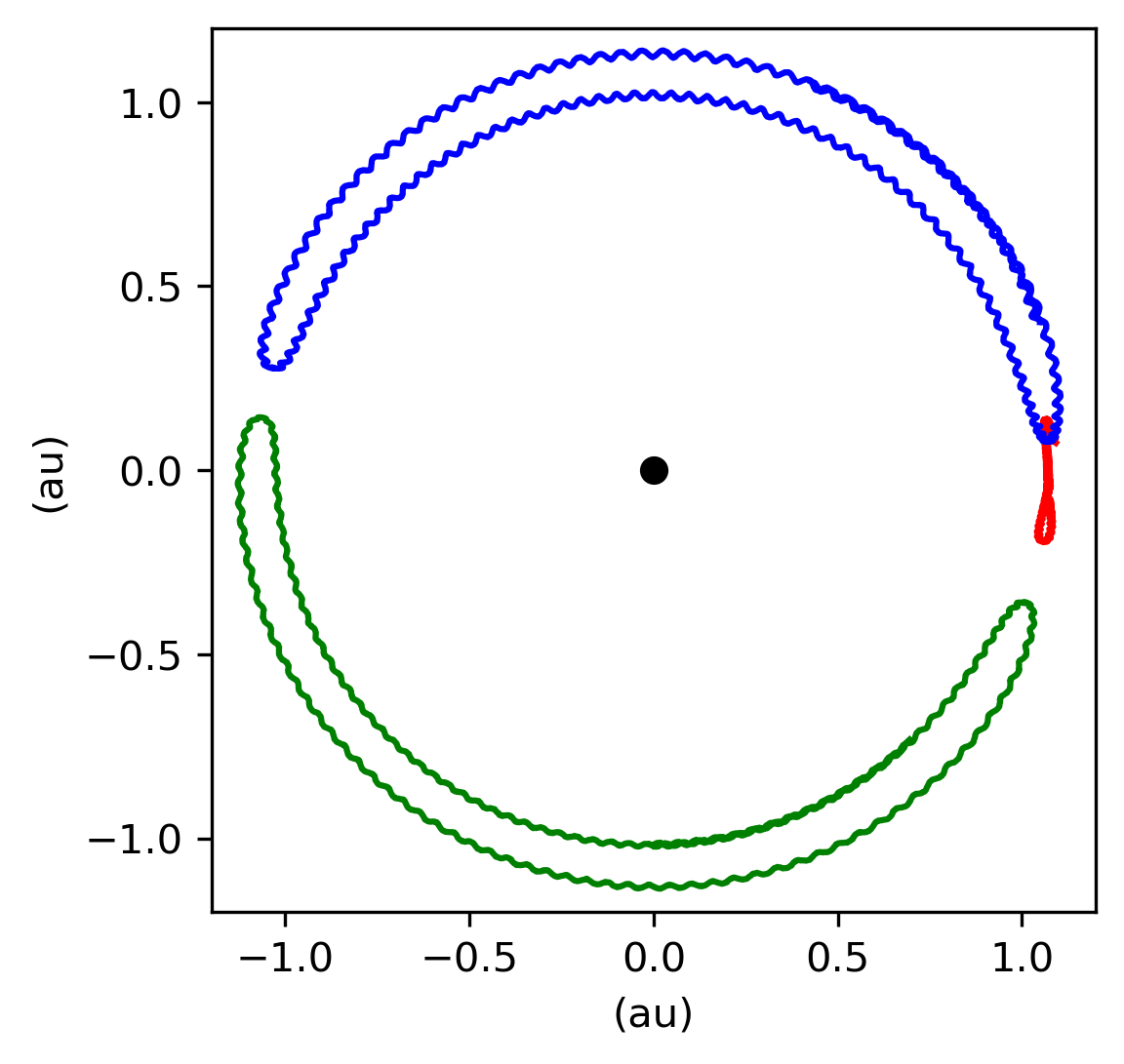}   	
   \includegraphics[width=0.66\columnwidth]{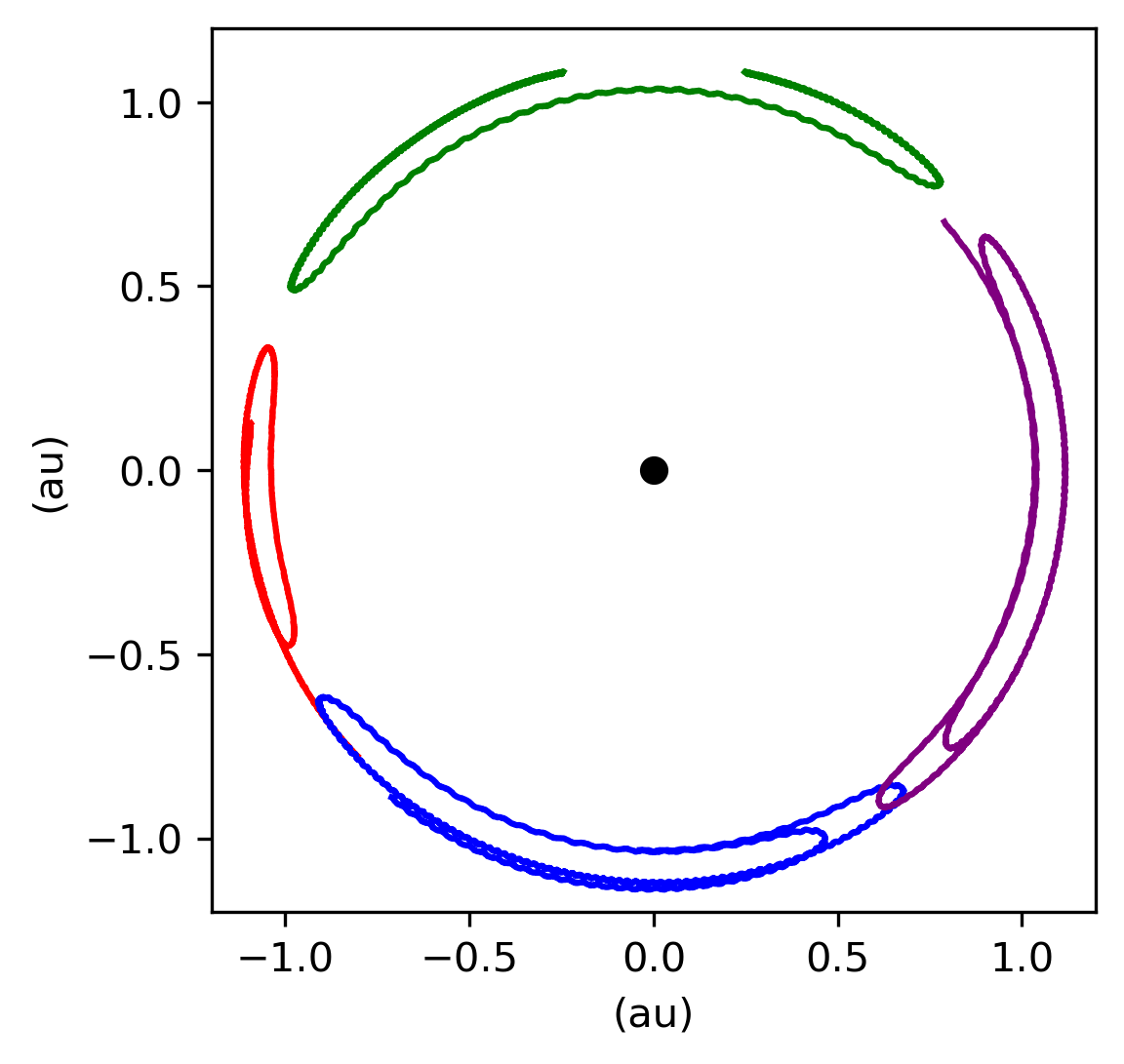} 
\\
  \includegraphics[width=0.66\columnwidth]{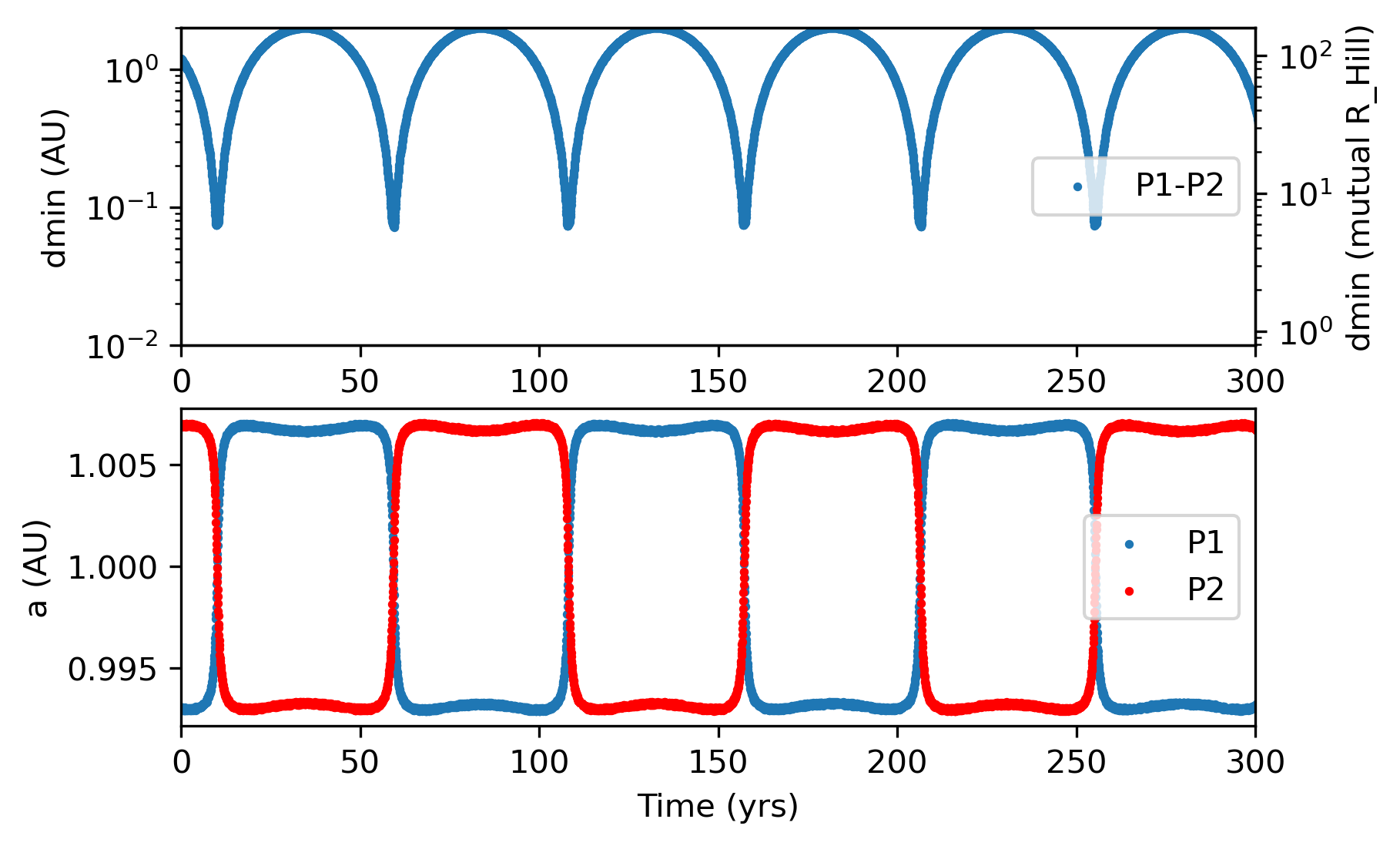}
	\includegraphics[width=0.66\columnwidth]{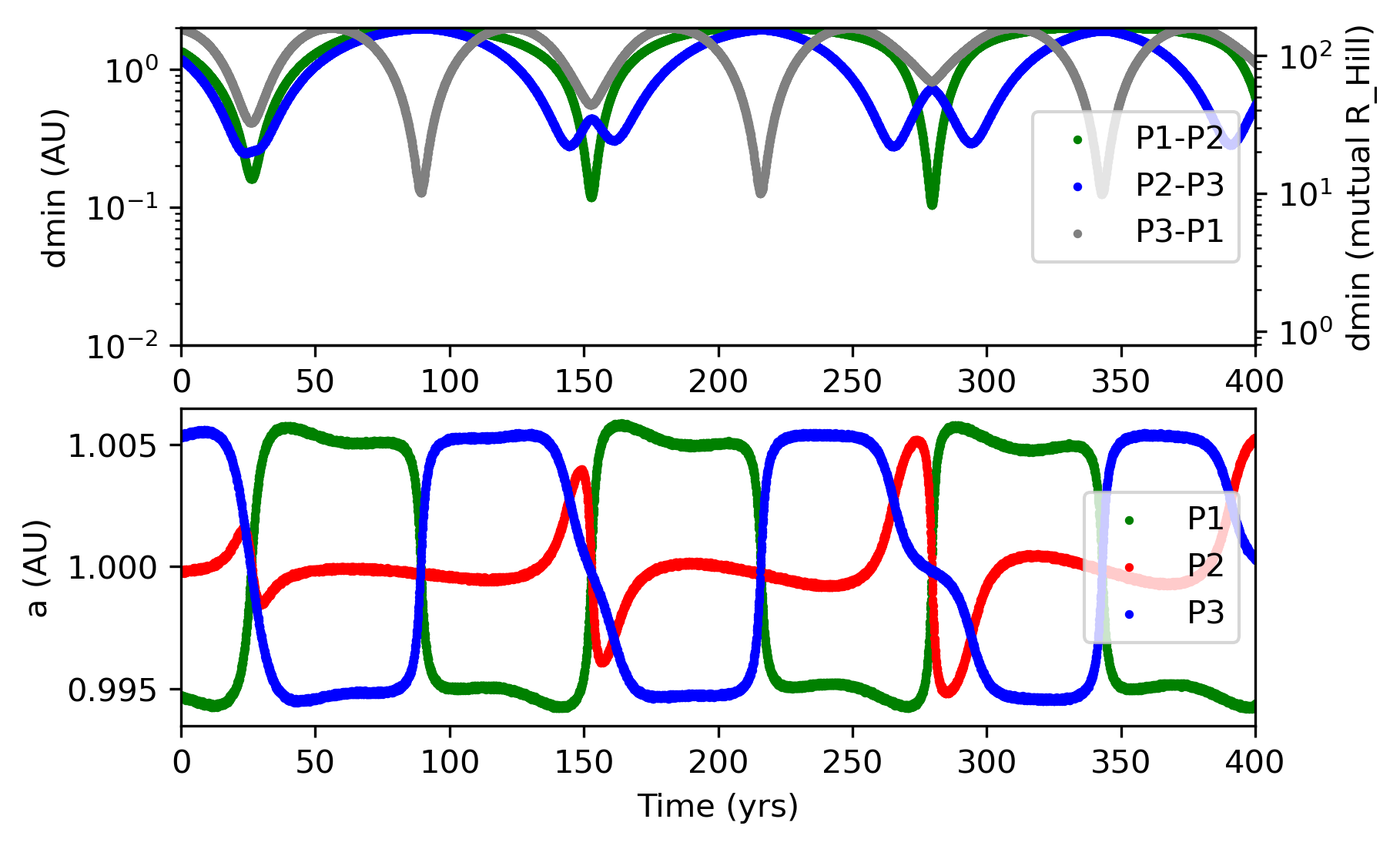}
 	\includegraphics[width=0.66\columnwidth]{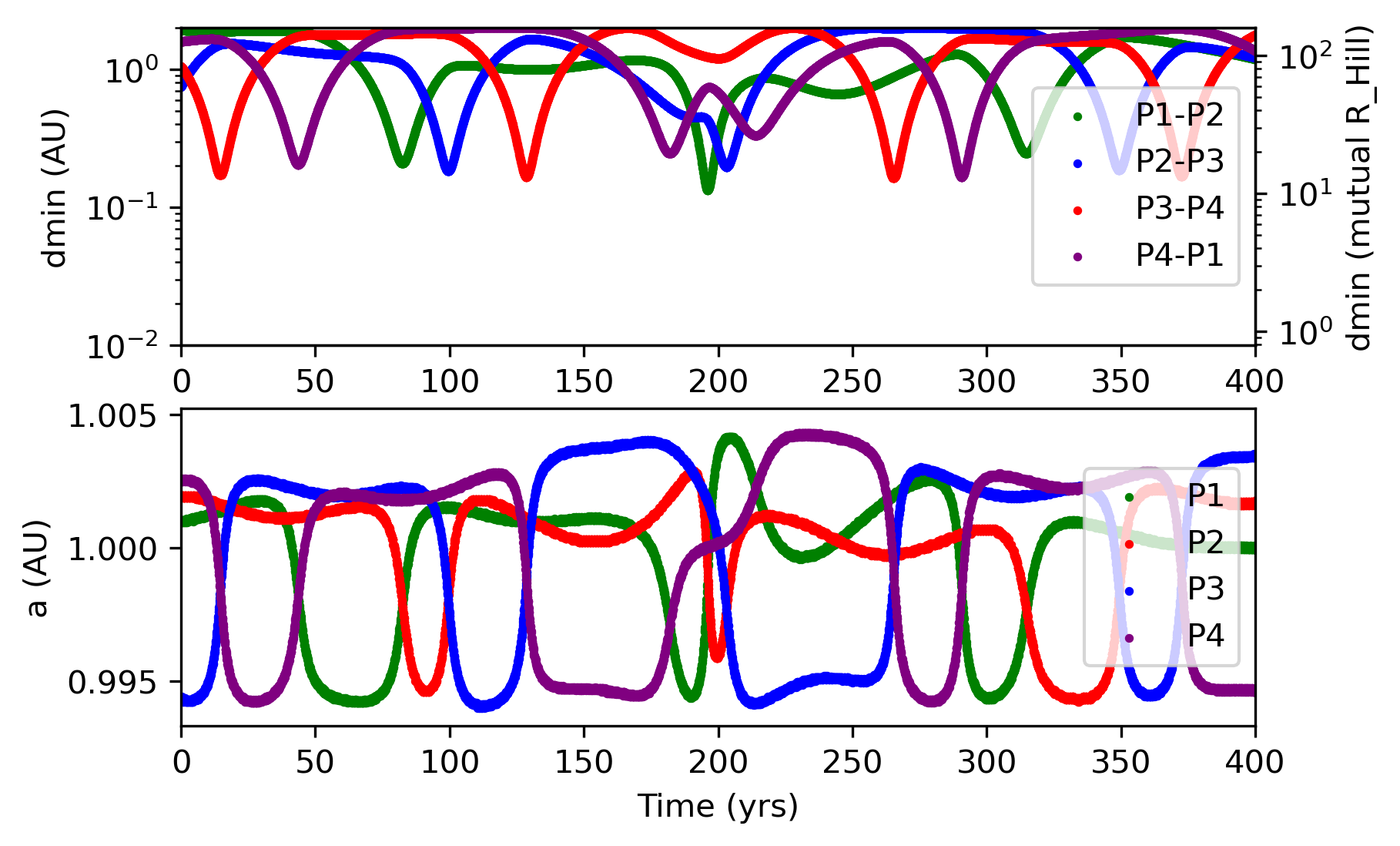}  
        \caption{Example of a 2- (left), 3- (center) and 4-planet (right) horseshoe systems.  In each case the planets are $1 \mearth$.  {\bf Top:} The libration of the planets seen in a co-moving frame, with the radial excursion of each planet expanded by a factor of 10 to improve visibility.  {\bf Bottom:} The distance between the two planets (measured in both au and mutual Hill radii) and their orbital semimajor axes, both as a function of time. During each close approach between the two planets, the exchange in energy causes an orbital flip in semimajor axis.  The small wobbles along some of the horseshoes are at the orbital frequency and come from misalignments in pericenter at the time of encounter~\citep[as shown by][]{dermott81b}.        }
    \label{fig:234pl}
\end{figure*}

\subsection{Survival}

Before discussing their dynamics, let us first consider the survival of co-orbital systems. It was shown in previous work that horseshoe orbits are only stable for primary-to-secondary mass ratios above 1200~\citep{laughlin02,cuk12b}.  As pointed out by \cite{cuk12b}, this is the reason that no horseshoe orbits have been found among Jupiter's Trojans~\citep{stacey08}.  In addition, \cite{cuk12b} showed that horseshoe orbits are typically destabilized by approaches smaller than ${\sim}5$ Hill radii.  In a system of equal-mass planets, this translates to a close approach limit of ${\sim}4$ mutual Hill radii.

Figure~\ref{fig:survtime} shows the survival time of simulations with 2 to 24 co-orbital Earth-mass planets as a function of their initial inter-planetary spacing.  The critical separation for Gyr stability of 2-planet systems is roughly 6 mutual Hill radii.  To be precise, all simulations with planets initially closer than 6 mutual Hill radii were unstable while all those initially farther apart than 6.75 were stable.  This stability limit appears to be roughly independent of planet mass for 2-planet systems (see Sections 2.4). 

Figure~\ref{fig:survtime} shows that the critical initial separation for long-term stability increases with planet mass, but only up to a limit. For 3-planet systems the critical stability limit is ${\sim}16$ mutual Hill radii, or roughly $12^\circ$ in angle along the orbit. For a larger number of planets the stability limit is roughly constant, lying at $19-21$ mutual Hill radii (or $\sim 15^\circ$) for the different sets of simulations.  Surprisingly, for the most closely-packed systems that we tested, with $N_p = 24$ planets, the stability limit was actually closer than for $N_p = 4-20$.  We attribute this to the horseshoe dynamics (that we will explore in the next Section): for $N_p = 20-24$, the azimuthal amplitude of horseshoe librations is small enough that close approaches remain more distant than for systems with fewer planets that must travel farther to encounter one another.

We tested the long-term dynamical stability of a subset of horseshoe constellations and found them to be stable for 10 Gyr. \cite{dermott81a} hypothesized that horseshoe orbits should often go unstable on long timescales.  Assuming a random walk in the phase space of horseshoe libration, they calculated a critical stability timescale 
\begin{equation}
    \tau_{crit} = T/\mu^{5/3},
\end{equation}
\noindent where $T$ is the orbital period and $\mu$ is the planet-to-star mass ratio.  For systems of Earth-mass planets, $\tau_{crit}$ is roughly 1 billion years, similar to the duration of our simulations.  We tested the very long-term stability of five selected simulations -- with horseshoe constellations containing 2, 3, 4, 6, and 12 planets -- for 10 Gyr.  All remained stable.

\subsection{Horseshoe dynamics}

Now we examine the dynamics of the systems themselves.  Inspection of the surviving systems from Fig.~\ref{fig:survtime} shows that the planets do not remain stationary with respect to each other, as would be the case for a ring of equally-spaced planets like those simulated by \cite{salo88} and \cite{smith10}.  Rather, our systems undergo horseshoe-type oscillations. 

Figure~\ref{fig:234pl} shows examples of 2-, 3- and 4-planet horseshoe systems from our simulations.\footnote{We have created animations showing the short-term movement of a selection of our systems (in inertial and co-rotating frames), which can be viewed at this url: \url{https://www.youtube.com/playlist?list=PLelMZVM3ka3F335LGLxkxrD1ieiLJYQ5N}.} The top panel for each system shows, in a frame that is co-moving at the mean mean motion, a top-down view of a short interval (generally ${\sim}100$ years) during which the planets have traced out a horseshoe-type shape. While the shapes vary, the evolution of each system follows a characteristic pattern: a close approach between planets (evident as a minimum in dmin in the Figure) triggers an exchange in orbital energy that causes a swap in orbital distance.  


The 2-planet system from Fig.~\ref{fig:234pl} is reminiscent of Janus and Epimetheus, as each planet traces out its own horseshoe between close approaches as viewed in a co-moving frame~\citep[see][for a nice discussion]{laughlin02}.  The orbital swaps are easy to identify in the 2-planet case, with each planet's semimajor axis flipping between roughly 0.0095 and 1.005 au every close approach. For the 3- and 4-planet systems the evolution is more complex, as each planet undergoes horseshoe oscillations with each of its neighbors. Yet the pattern of close approaches can be clearly connected with the planets' orbital evolution, and every close approach between planets A and B causes a flip in those planets' orbits. 

The dynamics of planets on horseshoe orbits is determined by their close approaches~\citep{dermott81a,murraydermott99}.  In the idealized, circular restricted three-body problem -- which includes a star, a massive planet, and a massless third body -- angular momentum and energy are not conserved, but the Jacobi Constant $C$ is~\citep[see][]{murraydermott99}. Joseph-Louis Lagrange showed that there are five equilibrium points co-orbiting with the planet; in addition to the L4 and L5 equilibrium points sixty degrees in front of and behind the planet's orbit, the equilibrium points L1, L2, and L3 lie on an imaginary line joining the two main masses of the system, and are dynamically unstable.  The dynamics of a co-orbital particle is determined by its Jacobi constant. $C$ is a minimum at the L4 and L5 points, higher at L3, and at a maximum at L1 and L2~\citep[see][]{dermott81a}.  A co-orbiting particle is on a tadpole orbit if $C_{L4,L5} < C < C_{L3}$ and on a horseshoe orbit if $C_{L3} < C < C_{L1,L2}$. Our setup does not strictly fall within the restricted three-body problem but the concept remains unchanged~\citep[see][for a Hamiltonian horseshoe formalism for two massive planets]{vok14}.  

\begin{figure}
 	\includegraphics[width=\columnwidth]{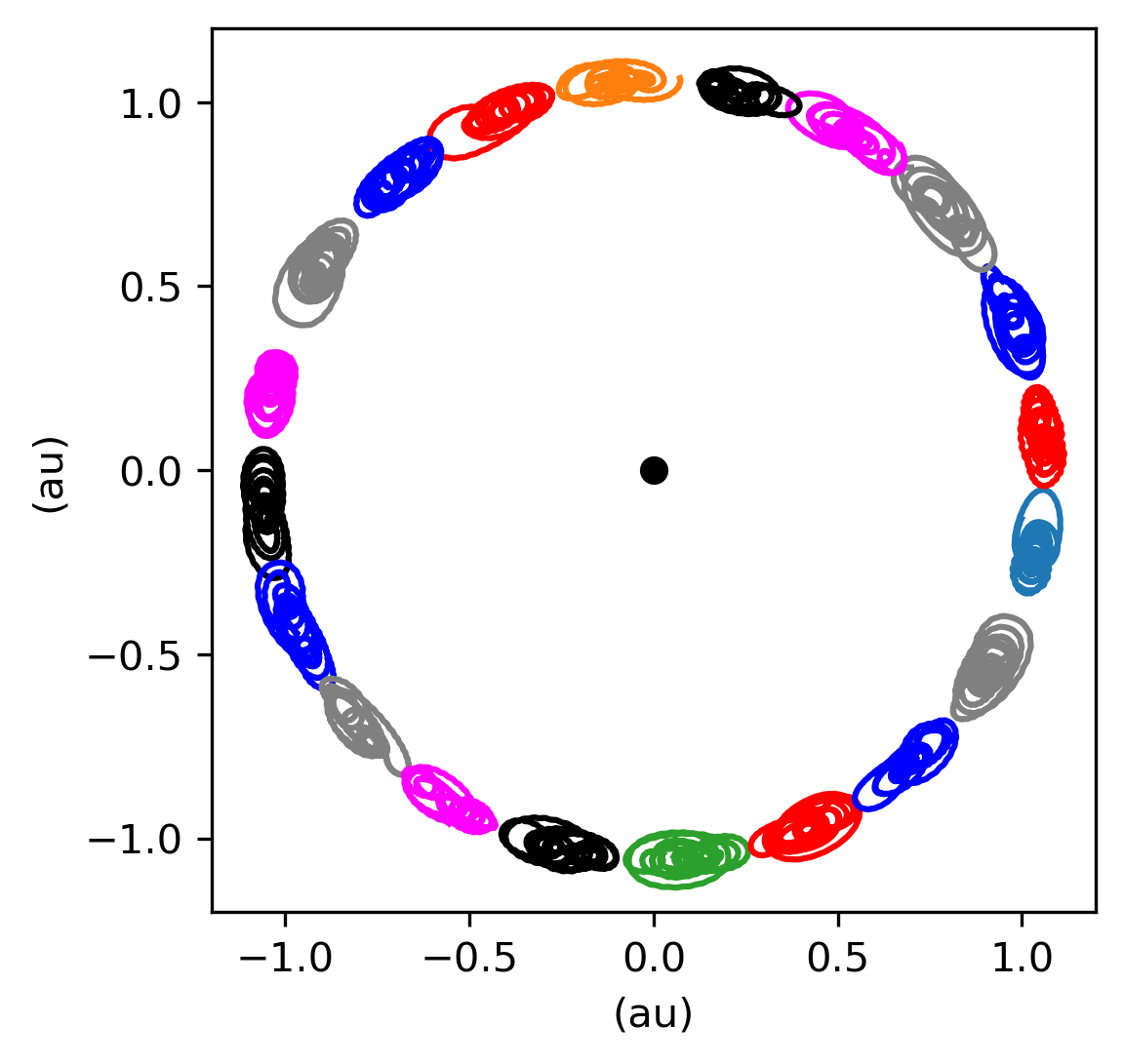}
\caption{Movement over roughly 275 years of a 20-planet horseshoe constellation (in a co-rotating frame). Each planet's libration is far more confined than in horseshoe systems with fewer planets (see Fig.~\ref{fig:234pl}).  As in Fig~\ref{fig:234pl}, the radial excursion of each planet was increased by a factor of 10.}
    \label{fig:20E}
\end{figure}

In our simulations, co-orbiting planets undergo horseshoe librations with their immediate neighbors.  The azimuthal extent of horseshoe libration depends on the number of co-orbiting planets, and, for more than two planets, may change in time (see Fig.~\ref{fig:234pl} for the case of 2, 3, and 4 planets). Figure~\ref{fig:20E} shows the horseshoe librations of a 20-Earth co-orbital system in which each planet's libration is so strongly confined that the horseshoe shape is barely apparent. In fact, the average separation between planets in this case is only $18^\circ$, such that neighboring planets are much closer than their mutual L4/L5 equilibrium points. The distinction between tadpoles and horseshoes in this regime becomes somewhat ambiguous.  Nonetheless, such systems can remain stable for Gyr timescales and longer (see Fig.~\ref{fig:survtime}).

\subsection{Effect of the planet masses and mass ratio}

\begin{figure*}
	\includegraphics[width=0.66\columnwidth]{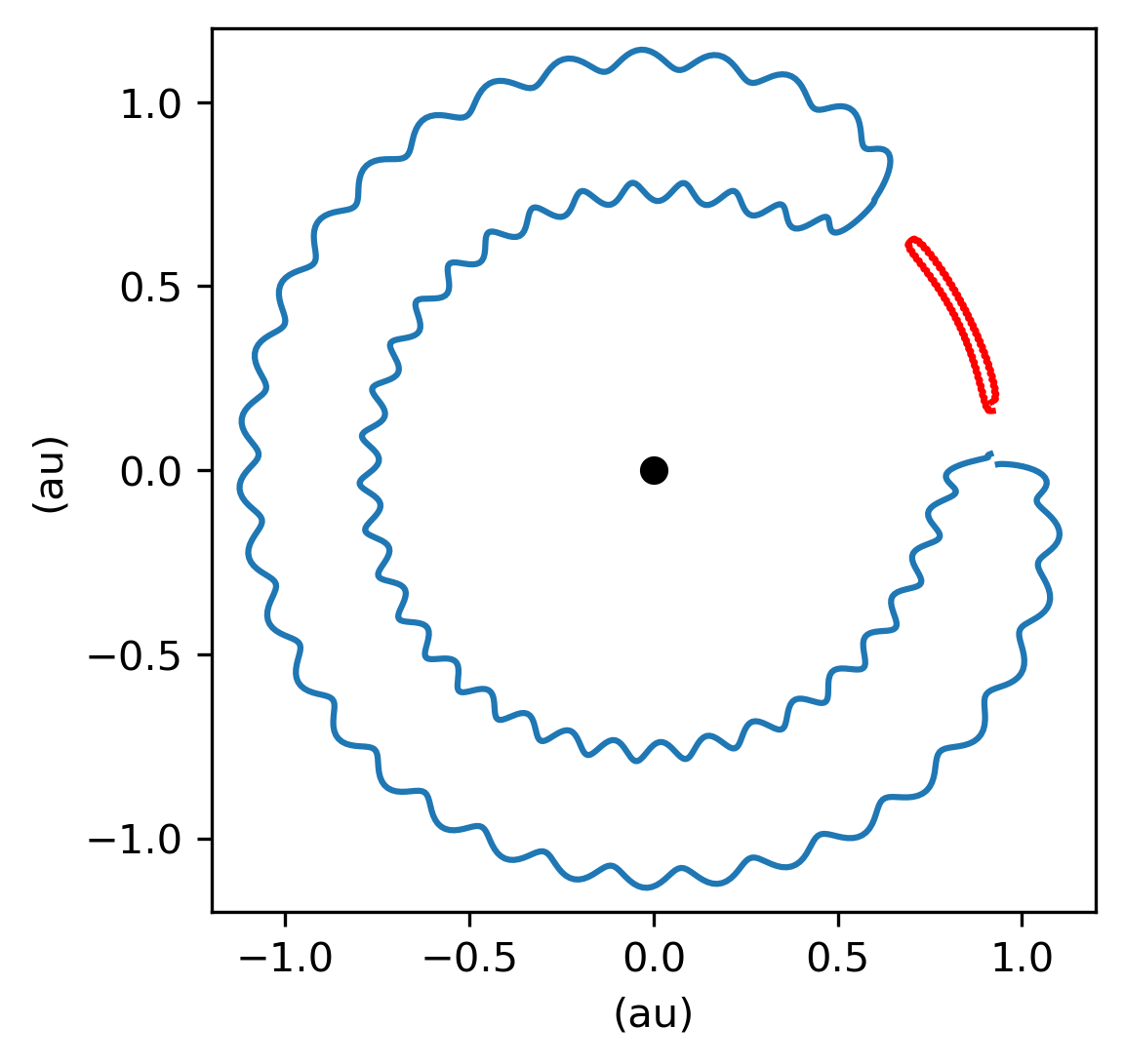}
 	\includegraphics[width=0.66\columnwidth]{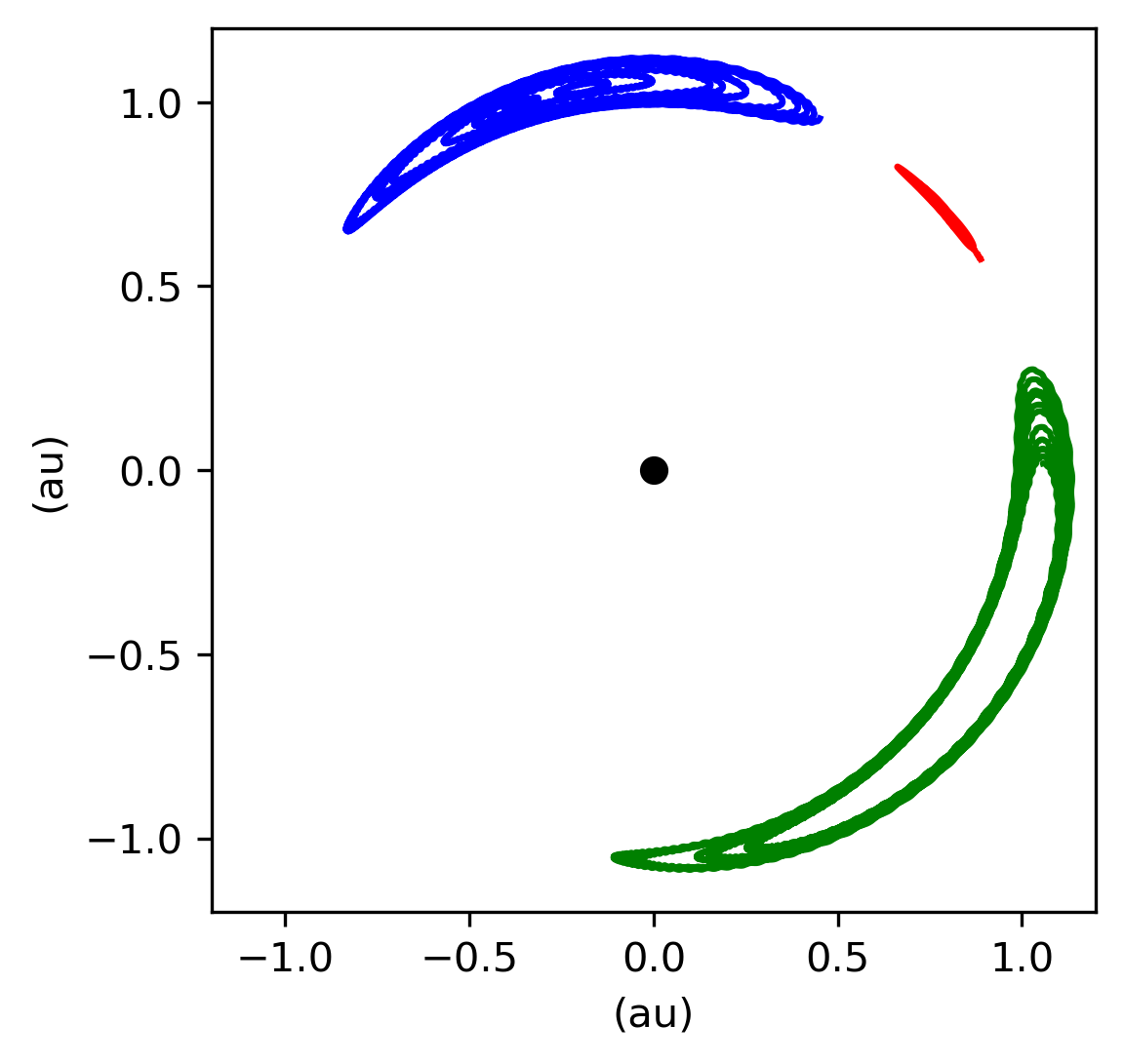}
  	\includegraphics[width=0.66\columnwidth]{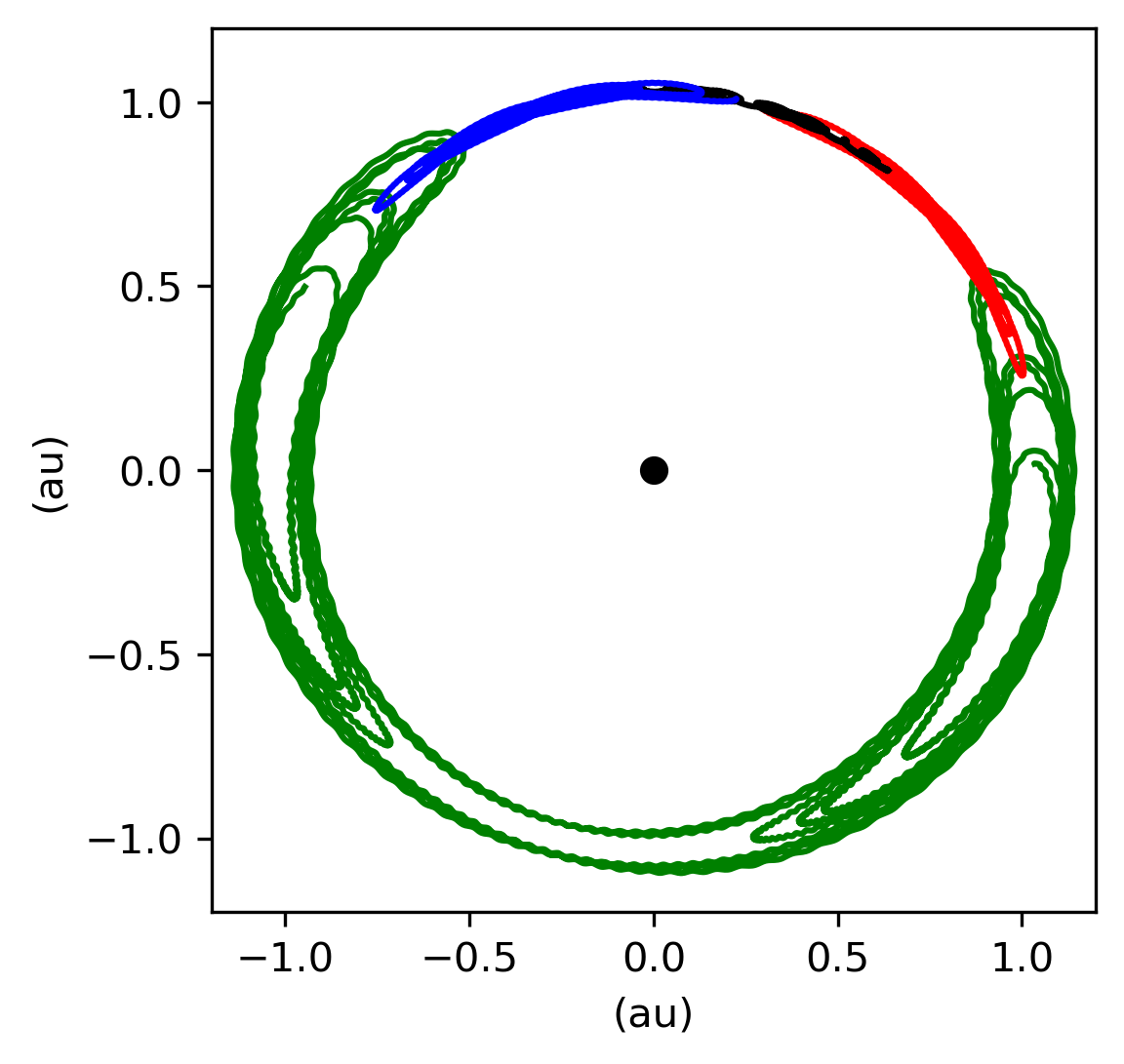}
\caption{Evolution of systems with unequal-mass planets.  {\bf Left:} A two-planet system with planets of $1 \mearth$ and $10 \mearth$ with initial interplanetary separation of $7 R_{HM}$.  {\bf Center:} A three-planet system with masses of, in order, $1 \mearth$, $10 \mearth$, and $1 \mearth$ with initial interplanetary separation of $25 R_{HM}$.  {\bf Right:} A four-planet system with masses of in order, $1 \mearth$, $10 \mearth$, $1 \mearth$, and $10 \mearth$, with initial interplanetary separation of $20 R_{HM}$. The two- and four-planet systems exhibit horseshoe dynamics but the three-planet case is only stable in a tadpole dynamical state, with each of the $1 \mearth$ planets librating about the Lagrange points of the $10 \mearth$ planet. As in Fig~\ref{fig:234pl}, the radial excursion of each planet was increased by a factor of 10.}
    \label{fig:rings}
\end{figure*}

We performed additional simulations to test the role of planet mass. For 2-planet systems, we tested systems with equal-mass planets of 1, 8, 64, and 512 Earth masses (the factor of 8 between successive tests was chosen because it doubles the Hill radius).  We found that for planet masses up to $64 \mearth$ the stability limit of ${\sim}6$ mutual Hill radii remained constant with planet mass (such that the critical separation in degrees increased as $M_p^{1/3}$). For planets of $512 \mearth$, the situation changes because none of our systems were in horseshoe libration. Rather, they followed tadpole orbits.  This is consistent with the analysis of \cite{dermott81a}, who showed that the parameter space available for horseshoe orbits decreases with increasing satellite mass. This also agrees with the results of \cite{laughlin02} and \cite{cuk12b}, who showed that horseshoe oscillations are unstable for two equal-mass planets above roughly a Saturn mass.

We also simulated selected systems with different-mass planets.  Figure~\ref{fig:rings} shows three examples of such systems, with 2, 3, and 4 planets, respectively.  The first system, with two planets with masses of 1 and $10 \mearth$, undergoes simple horseshoe libration reminiscent of Janus and Epimetheus.  The second system, with three planets with masses of 1, 10, and $1 \mearth$, behaves differently.  The lower-mass planets each are on tadpole orbits, librating about the L4/L5 points of the more massive central planet.  The third example contains planets of 1, 10, 1, and $10 \mearth$.  Its evolution contains both tadpole and horseshoe elements.  One of the $1 \mearth$ remains trapped between the two $10 \mearth$ planets and the three are on mutual tadpole orbits.  Yet the other $1 \mearth$ planet is on a horseshoe orbit, undergoing repeated close encounters with each of the $10 \mearth$ planets.  

We did not systematically vary the number of planets in systems with different-mass planets, as preliminary simulations found that the parameter space was largely unstable (a limited exploration is presented in Section 2.8). 

\subsection{Dynamical heating and disruption}

We performed a few simulations with high-frequency outputs to explore exactly how horseshoe systems break. We focus our attention on the case of a 3-planet horseshoe system in which the planets were initially separated by 15.4 mutual Hill radii and that became unstable after 53.7 Myr of integration.

\begin{figure*}
	\includegraphics[width=\columnwidth]{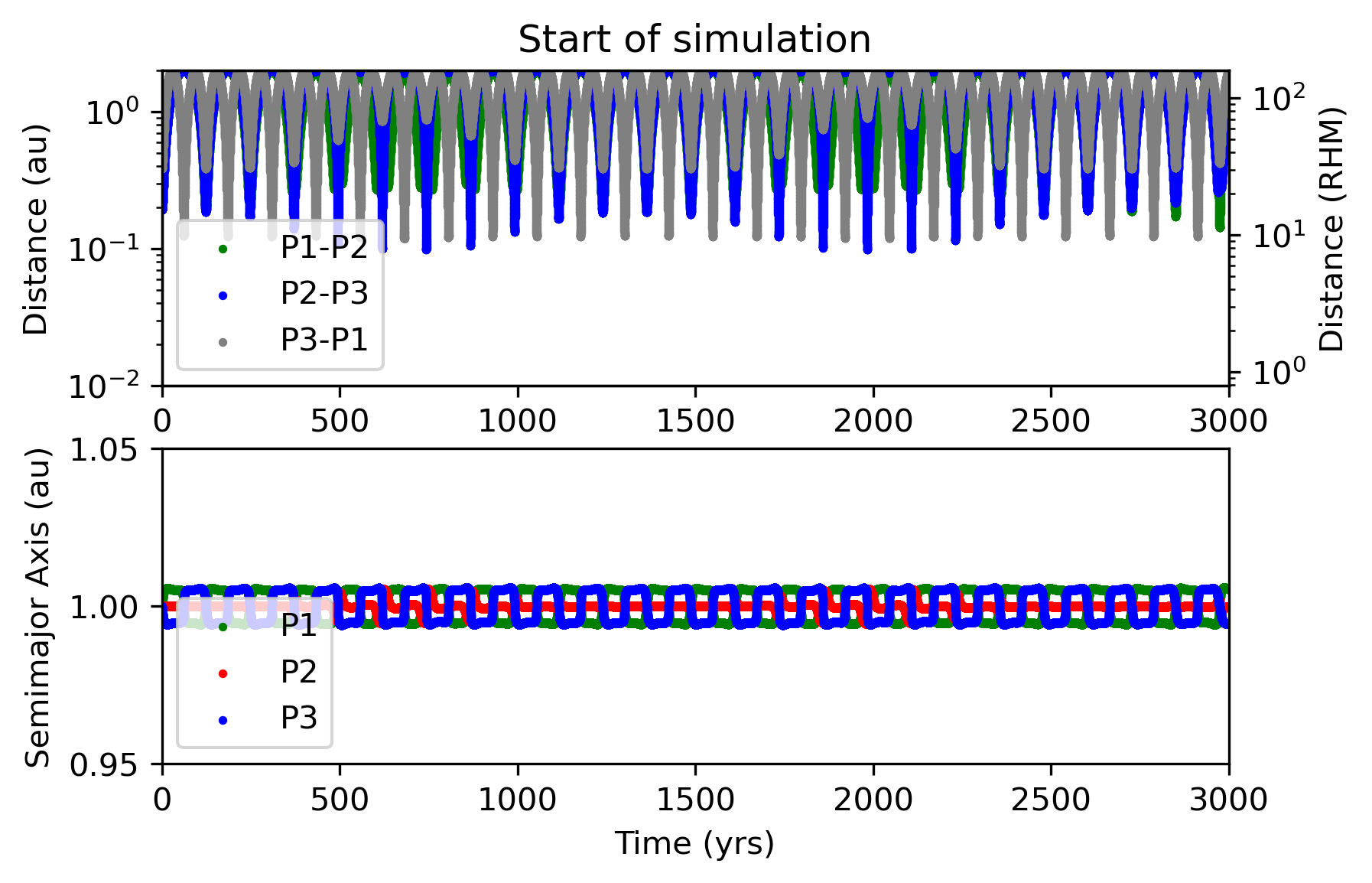}
 	\includegraphics[width=\columnwidth]{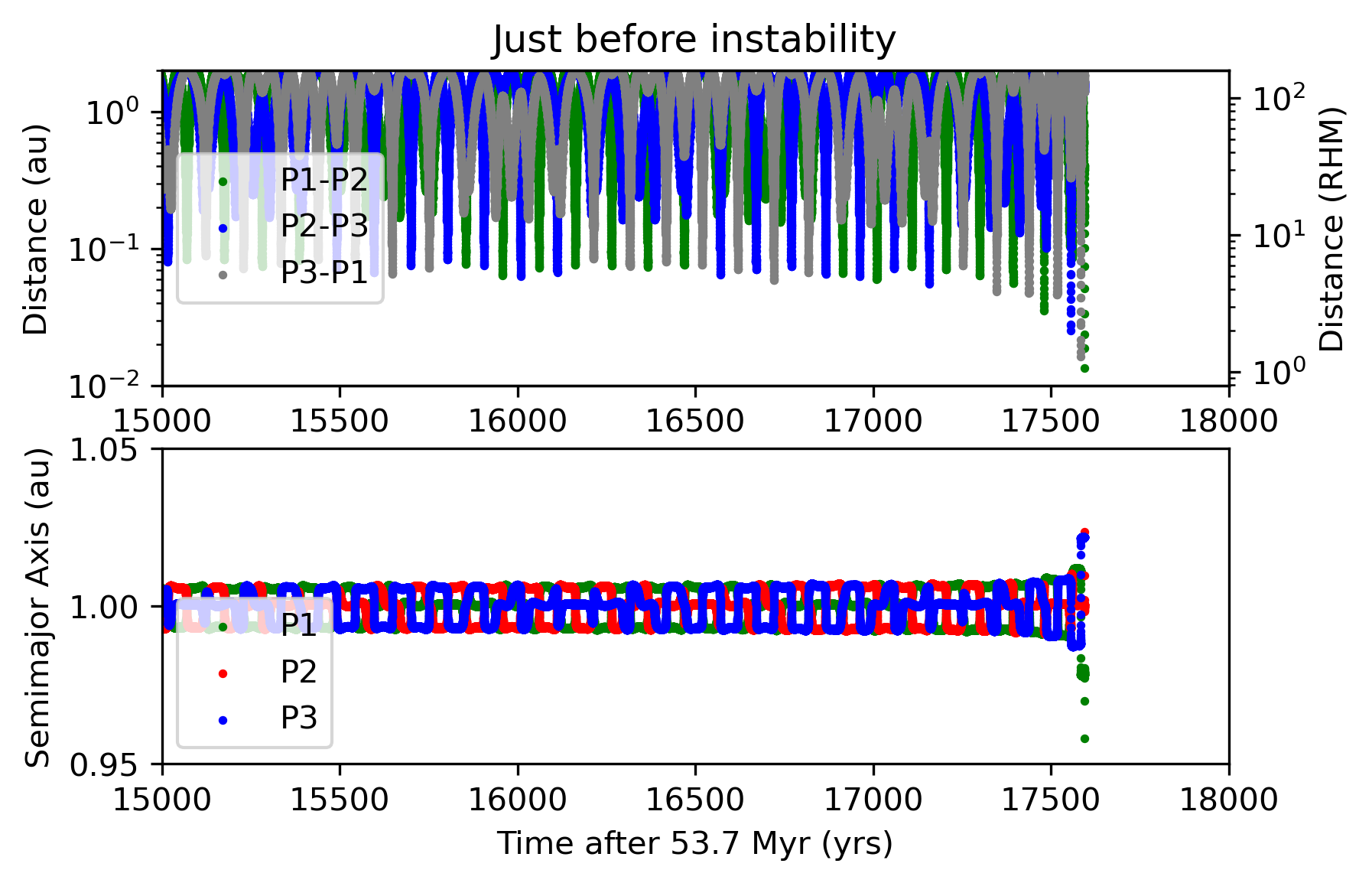}
    \caption{Dynamical heating and disruption of a 3-planet horseshoe system. The left panel shows a 3000 year snapshot of the system immediately after the start of the simulation, and the right panel shows a similar snapshot immediately prior to destabilization roughly 53.7 Myr later.  For both cases, the top panel shows the distances between each pair of planets (in both au and mutual Hill radii) as a function of time, and the bottom shows the orbital semimajor axis of each planet in time.}
    \label{fig:dyn_heating}
\end{figure*}

Figure~\ref{fig:dyn_heating} shows two snapshots in the evolution of the system, first immediately after the start of the simulation and then just before its demise.  There are three key differences between the two snapshots: just before the instability, the horseshoe libration amplitudes were larger, the libration frequency was higher, and the close approaches between planets were closer.  These three factors are interrelated, as closer approaches between planets imply an increase in exchange of orbital energy and therefore a larger radial horseshoe amplitude, which implies a shorter time before the next close approach.  Eventually, the close encounters crossed the limit for horseshoe stability, shown by \cite{cuk12b} to be ${\sim}5$ Hill radii, which translates to ${\sim}4$ mutual Hill radii for equal-mass planets.

\subsection{Survival during stellar post main sequence evolution}

We tested the long-term stability of horseshoe constellation systems in Section 2.2, but what about their survival in the face of the long-term evolution of their host stars? To address this question, we provide an analytical argument, and then reinforce it through $N$-body integrations.

When host stars leave the main sequence, they typically lose 50-80 per cent of their mass through stellar winds during the giant branch phases of stellar evolution. With regard to the stability of horseshoe constellation systems, this mass loss creates competing effects by expanding the orbits of the planets but also increasing the size of their Hill spheres. Both effects may quantified by considering the initial mass of the star along the main-sequence, $M_{\star}^{(\rm MS)}$ and the final mass of the star, after it has become a white dwarf, $M_{\star}^{(\rm WD)}$.

Our goal is to assess whether horseshoe constellations will remain dynamically stable during their central star's post-main-sequence evolution. First, consider two adjacent planets in a horseshoe constellation that are separated by a true longitude difference of $\Delta \theta^{(\rm MS)}$. The two planets are in danger of becoming unstable if $\Delta \theta^{(\rm MS)}$ decreases as the Sun transitions into a white dwarf.  The pertinent question is, therefore, what is the maximum value of $\Delta \theta^{(\rm MS)}$ for which the distance between Hill spheres {\em decreases} due to post-main-sequence evolution?  To answer the question, we need to compare the distance between the planets' Hill spheres during the white phase with those during the main sequence. 

Start with the distance between the two planets themselves, $r$. Along both phases of stellar evolution,

\begin{equation}
r^{(\rm MS)} = 2 a^{(\rm MS)} \sin{\left(\frac{\Delta \theta^{(\rm MS)}}{2}\right)},
\end{equation}

\begin{equation}
r^{(\rm WD)} = 2 a^{(\rm WD)} \sin{\left(\frac{\Delta \theta^{(\rm WD)}}{2}\right)},
\end{equation}

\noindent{}here $a$ refers to semi-major axis. As long as the planets are within several hundred au of the star, mass loss will not change the relative true longitude of the planets, and increase their semimajor axes in inverse proportion to the mass loss \citep{veras11}. Hence,

\begin{equation}
\frac{r^{(\rm WD)}}{r^{(\rm MS)}} =   
\frac{a^{(\rm WD)}}{a^{(\rm MS)}} \approx
\frac{M_{\star}^{(\rm MS)}}{M_{\star}^{(\rm WD)}}.
\end{equation}

\noindent{}The Hill sphere radius change for a single planet has a different functional dependence on mass loss, and is given by \citep{payne2016}

\begin{equation}
\frac{R_{\rm H}^{(\rm WD)}}{R_{\rm H}^{(\rm MS)}} \approx
\left(\frac{M_{\star}^{(\rm MS)}}{M_{\star}^{(\rm WD)}}\right)^{4/3}.
\end{equation}

The condition for the distance between Hill spheres to decrease is given by 

\begin{equation}
\frac{r^{(\rm WD)} - 2 R_{\rm H}^{(\rm WD)}}
     {r^{(\rm MS)} - 2 R_{\rm H}^{(\rm MS)}} < 1,
\end{equation}

\noindent{}which becomes

\begin{equation}
{\rm max}\left(\Delta\theta^{(\rm MS)}\right)
=
2
\sin^{-1}\left\lbrace
\left(\frac{m_1}{3M_{\star}^{(\rm MS)}}\right)^{1/3}
\left[
\frac
{\left(\frac{M_{\star}^{(\rm MS)}}{M_{\star}^{(\rm WD)}}\right)^{4/3} - 1}
{\frac{M_{\star}^{(\rm MS)}}{M_{\star}^{(\rm WD)}}-1
}
\right]
\right\rbrace
\nonumber
\end{equation}

\begin{equation}
\ \ \ \ \ \ \ \ \  \approx 2 \sin^{-1}
      \left[1.5-1.9 \times 10^{-2} \,{\rm rad}
      \left( \frac{m_1}{M_{\oplus}} \right)^{1/3}
      \left( \frac{M_{\star}^{(\rm MS)}}{M_{\odot}} \right)^{-1/3}
      \right]  
\nonumber
\end{equation}

\begin{equation}
\ \ \ \ \ \ \ \ \  \approx 1.7^{\circ}-2.2^{\circ}
      \left( \frac{m_1}{M_{\oplus}} \right)^{1/3}
      \left( \frac{M_{\star}^{(\rm MS)}}{M_{\odot}} \right)^{-1/3}.
\label{DegCrit}
\end{equation}

\noindent{}The range given in equation (\ref{DegCrit}) covers the entire range of stellar masses for which a single star can become a white dwarf. Further, note that equation (\ref{DegCrit}) is independent of $a^{(\rm MS)}$.

The equation indicates that for planets separated by more than about 2 degrees along a horseshoe orbit, their stability should {\it increase} during post-main-sequence evolution. This limit suggests that the stability of every system that was simulated along the main-sequence in Fig. \ref{fig:survtime} should increase during post-main-sequence stellar mass loss.

In order to support our conclusion, we conducted $N$-body simulations which include stellar evolution. We made use of the code from \cite{mustill18}, which includes the stellar evolution model from \cite{hurley00} and is built on the {\tt RADAU} integrator included in the {\tt Mercury} integration package~\citep{chambers99}. We modelled the evolution of a $1.0M_{\odot}^{(\rm MS)}$ star for between 1-8 Gyr after it left the main sequence, depending on the number of planets, for numerical reasons\footnote{Stability occurrences may be more naturally determined as a function of the the number of orbits instead of an absolute timescale. However, the number of planetary orbits corresponding to a fixed timescale decreases as the star loses mass. Therefore, for a fixed value of $a^{(\rm MS)}$ and a fixed simulation duration, the number of orbits completed would vary depending on the prescription used for stellar evolution.}.

We simulated a variety of different systems. The first set includes the following number of planets, all with $\Delta \theta^{(\rm MS)} = 20^{\circ}$ and $a^{(\rm MS)} = 5$~au: 2, 3, 4, 9, 12, and 18. These were run for 8 Gyr and all remained stable. We then repeated these simulations, again for 8 Gyr, but with $a^{(\rm MS)} = 2$~au. These also remained stable. Finally, we simulated a system with 36 planets, $\Delta \theta^{(\rm MS)} = 10^{\circ}$ and $a^{(\rm MS)} = 2$~au for 1 Gyr. Again, this system remained stable. These results all support our analytical conclusion.

\begin{figure}
	\includegraphics[width=\columnwidth]{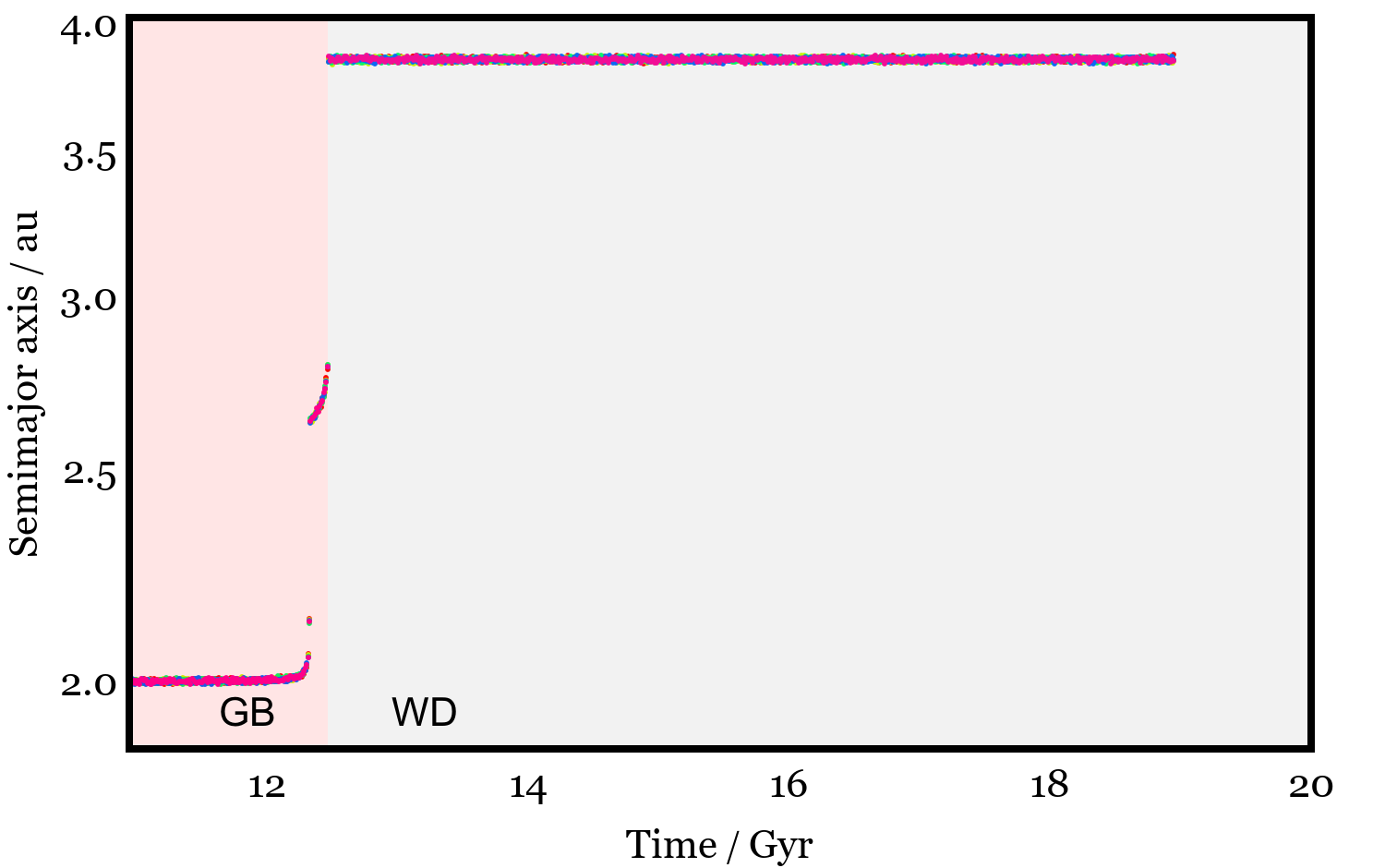}
    \caption{Survival of a multi-planet horseshoe system through its host star's post-main sequence evolution.  The planets' orbits expand in concert and retain their horseshoe configuration.  Here, the time is relative to the formation of the Sun-like central star.  The giant branch ('GB') and white dwarf ('WD') phases are labeled, and the short interval in between is the asymptotic giant branch phase.  We note that, while this particular simulation was for a horseshoe constellation of 9 planets, an analogous plot for a different number of planets would look identical. For a review of the effect of post-main sequence stellar evolution on the orbits of planets and small bodies, see \citet{veras16}.    }
    \label{fig:postms}
\end{figure}

We illustrate the orbital effect of post-main-sequence evolution on cohorts in Figure~\ref{fig:postms}. This schematic illustrates two different giant branch phases: the red giant phase and asymptotic giant branch phase. The mass loss profiles and timescales are different for each. However, the conclusion from equation (\ref{DegCrit}) is independent of these profiles. The systems maintain their horseshoe configuration because mass loss does not change their relative true longitudes.

\subsection{Multiple horseshoe rings}

Systems with more than one horseshoe ring can be stable if they are sufficiently separated.  To quantify this, we tested the stability limits of three rings of co-orbital planets, each containing a three-planet horseshoe, and compared it with standard systems with a single planet per orbit, for a planet mass of $1 \mearth$ and $3 \mearth$ planets.  For the horseshoe systems, we chose an initial separation of $20^\circ$ of the planets along each orbit, which corresponds to more than 25 mutual Hill radii, to ensure that each ring would be stable on its own (see Fig.~\ref{fig:survtime}). The only thing that varied between simulations was the separation between orbits.

\begin{figure}
	\includegraphics[width=\columnwidth]{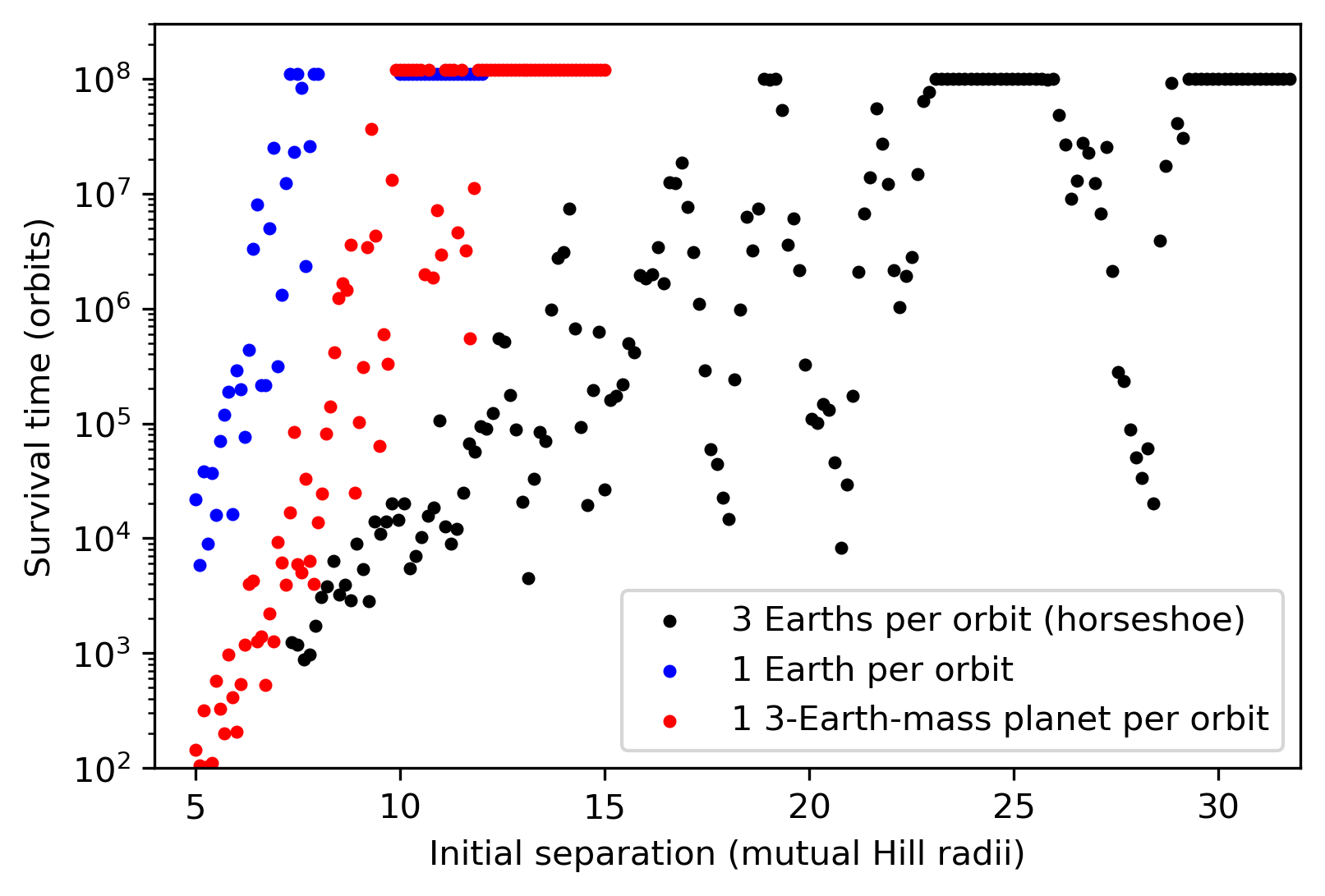}
    \caption{Survival of three different types of systems, each with three different orbits: systems with a single $1 \mearth$ planet per orbit, those with a single $3 \mearth$ planet per orbit, and systems with three $1 \mearth$ planets undergoing horseshoe libration.  The initial separation on the x axis is measured in units of mutual Hill radii between the two planetary orbits, assuming a $1 \mearth$ planet on each orbit. }
    \label{fig:3rings}
\end{figure}

\begin{figure*}
	\includegraphics[width=\columnwidth]{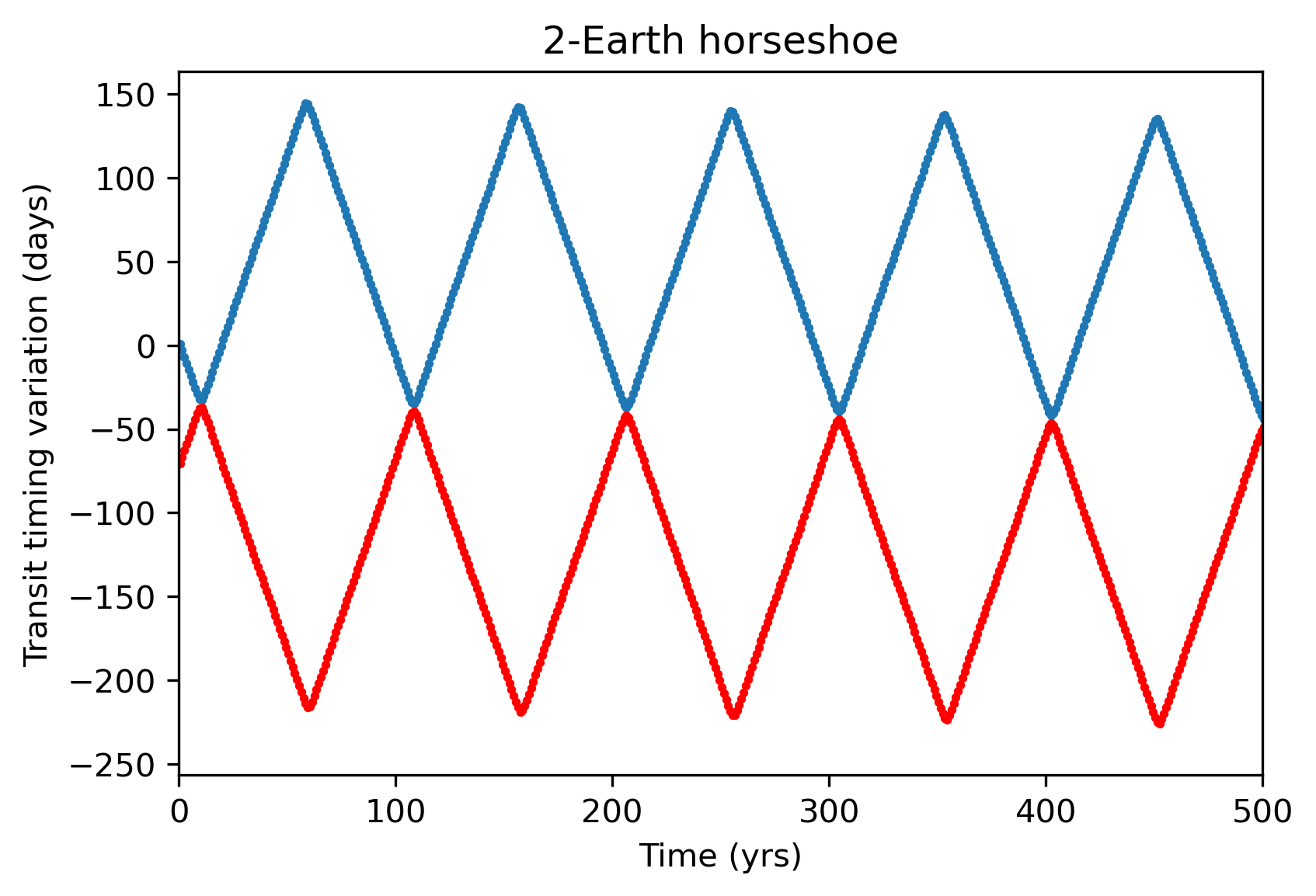}
 	\includegraphics[width=\columnwidth]{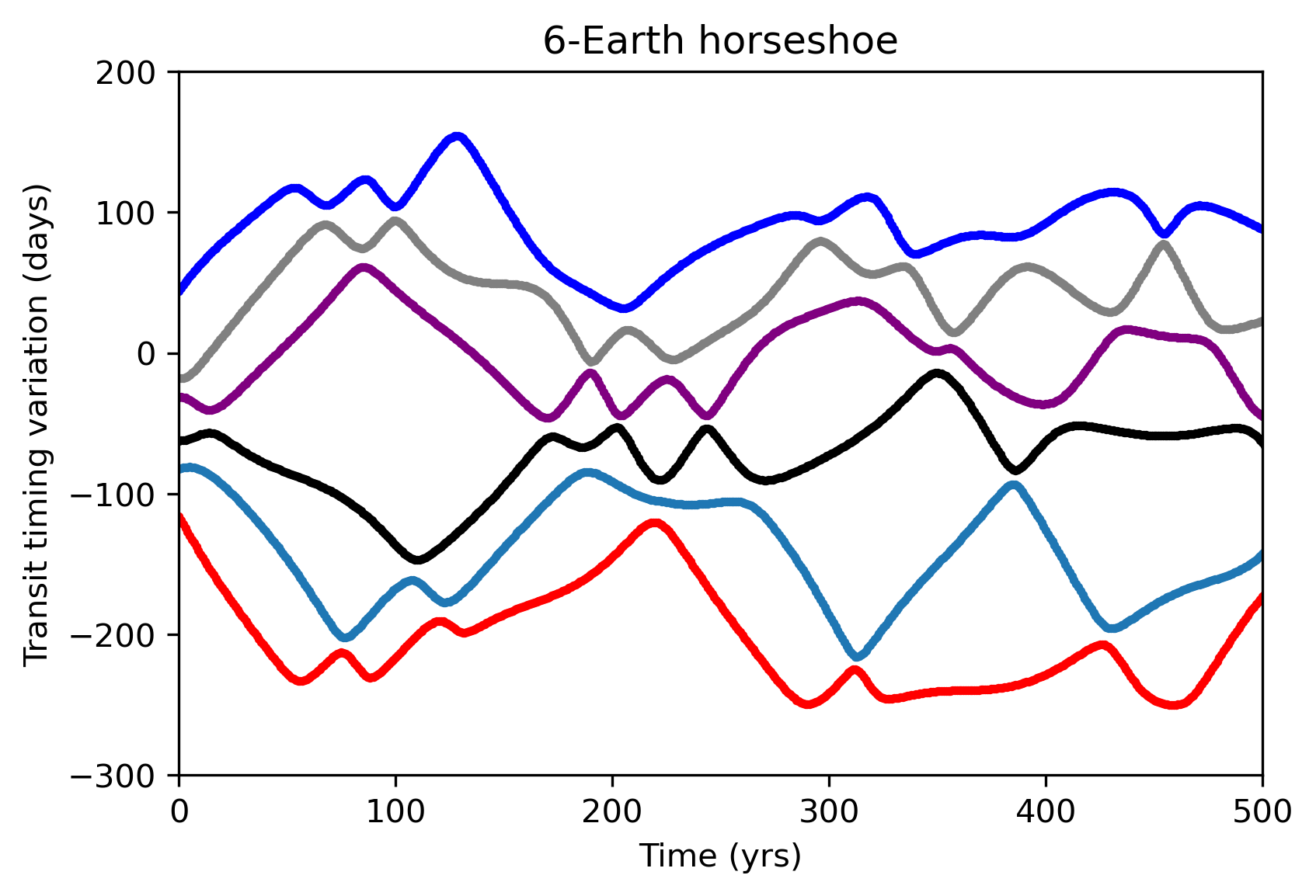} 
\caption{Illustrative examples of transit-timing variations of horseshoe systems with 2 (left) and 6 (right) planets. The y axis measures the deviation of the timing of each transit compared to when it would have occurred if each planet simply followed its average mean motion around the star (at a rate of $360^\circ \, \mathrm{yr^{-1}}$.) }
    \label{fig:TTV}
\end{figure*}

Figure~\ref{fig:3rings} shows the survival time of these different systems as a function of initial separation (in units of mutual Hill radii between adjacent orbits, assuming one planet of $1 \mearth$ per orbit).  By comparing the two systems with one planet per orbit, we recover the well-known result that lower-mass planets remain stable with more compact orbital configurations~\citep[e.g.][]{chambers96}.  In contrast, 3-planet systems are only stable if they are much more widely-spaced.  For scale, for an inner orbit at 1 au, a one-planet-per-orbit system was stable when the second orbit was wider than 1.10 au for $1 \mearth$ planets or 1.13 au for $3 \mearth$ planets.  In contrast, the first stable 'island' for 3-planet horseshoe systems (located at separations of $\sim22$-25 mutual Hill radii in Fig.~\ref{fig:3rings}) consists of system in which the second orbit is located beyond 1.34 au, and the more reliable stable region (beyond ${\sim}30$ mutual Hill radii in Fig.~\ref{fig:3rings}) included systems in which the second orbit was beyond 1.45 au.  Thus, even though 3-planet horseshoe systems have the same mass as systems with one $3 \mearth$ planet per orbit, rings of horseshoe systems are far less stable.  We attribute the instability of multiple-ring horseshoe systems to the vastly increased number of gravitational encounters between planets that may increase the likelihood of secular chaos. Among unstable systems with 3 co-orbital rings of 3 Earths, instability was triggered more than twice as often in the middle or outer rings than in the inner ring.

\subsection{Exotic configurations}

Out of curiosity we tested the stability of some exotic configurations of horseshoe constellations. 

As a first simple test, we asked the question of whether additional planets in horseshoe configurations would remain stable in the Solar System.\footnote{A number of studies have tested the stability of co-orbital asteroids of the planets~\citep[e.g.][]{erdi96,nesvorny02,scholl05}. Simulations show that co-orbitals of the terrestrial planets are usually not long-term stable due to perturbations from the other planets~\citep[e.g.][]{brasser02,scholl05}. } To test this, we started from the present-day orbits of all 8 Solar System planets, and then simple added additional co-orbital planets.  

Remarkably, we found that a number of configurations were stable for at least 1 Gyr.  For instance, a 3-Earth Solar System -- created by simply introducing 1-2 extra Earth-mass planets along Earth's orbit -- remained perfectly stable.  Likewise, the Solar System remained stable when we added one or two additional Venus-mass planets along Venus' orbit, one additional Mars along Mars' orbit, or one additional Mercury along Mercury's orbit.  No simulations were stable with co-orbitals of all four terrestrial planets at once.  However, in one simulation, the Solar System with three Earths and three Venuses remained stable.  Systems with larger number of horseshoe planets were not stable, nor were equal-mass co-orbitals of any of the giant planets.  

We also tried to determine how many planets could fit in a horseshoe ring including planets with different masses. In our preliminary tests shown in Section 2.4 we included four planets, alternating in mass between $1 \mearth$ and $10 \mearth$.  Using the same pattern of masses, we found that a 6-planet horseshoe system was stable, but 8- or 10-planet horseshoe systems were not.

\section{Transit timing variations of horseshoe constellation systems}

Transit-timing variations (TTVs) of co-orbital planets have been explored by several previous papers~\citep{ford06b,ford07,haghighipour13,vok14,veras16b}. In horseshoe systems, the transit timing variations signal of each planet will have an amplitude that matches the amplitude (in phase) of the horseshoe region it traces out, and a timescale related to the horseshoe libration period. \cite{vok14} studied this for 2-planet horseshoe systems, and showed that the transit-timing variations signal is triangle-shaped.  For adequate observations of both planets, one can deduce the mass ratio between the two planets from the relative TTV amplitudes.  In addition, information regarding the total planet mass- to star ratio can be obtained by time separation of the transits of the two planets~\citep{vok14}.

\begin{figure*}
	\includegraphics[width=\columnwidth]{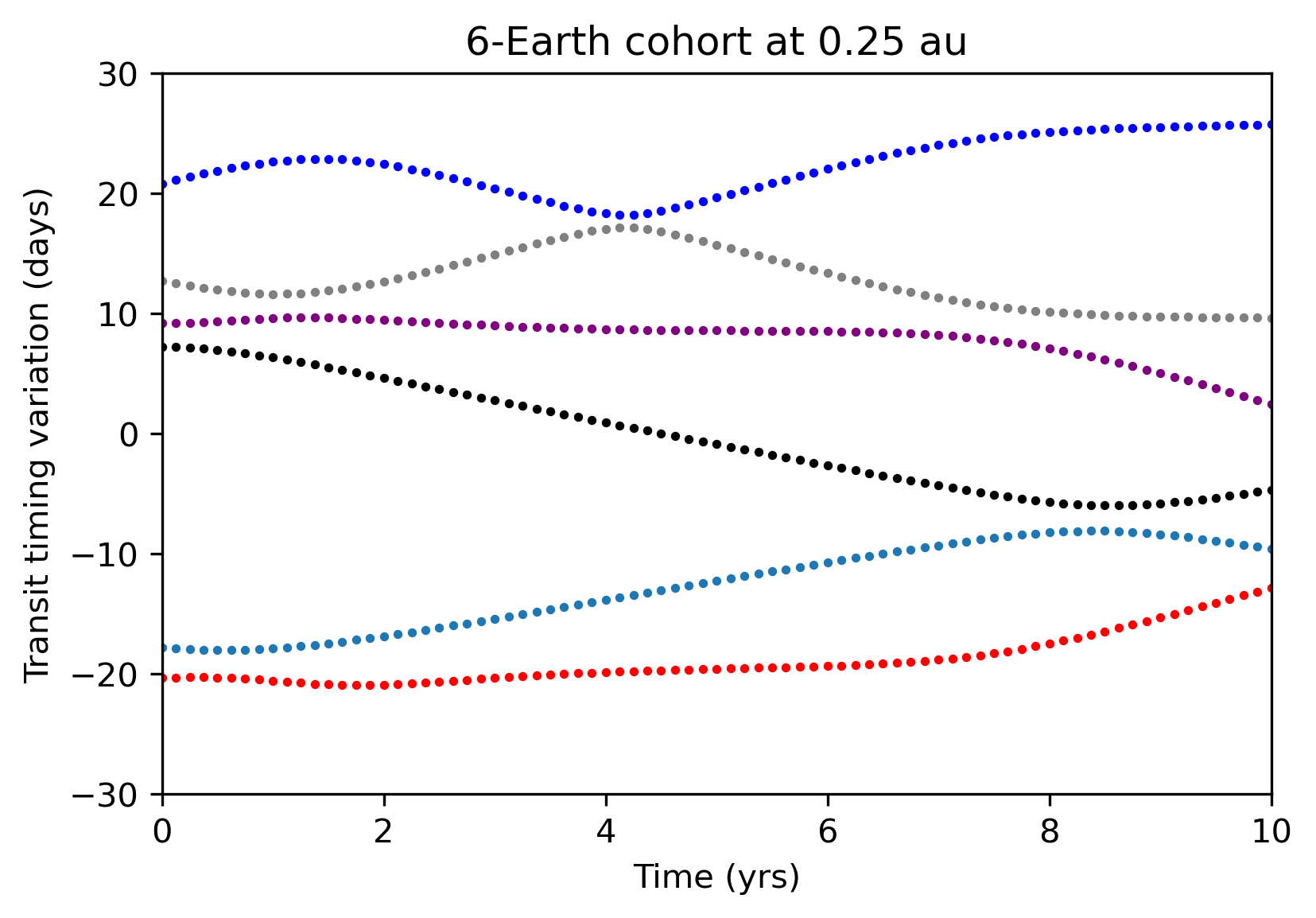}
 	\includegraphics[width=\columnwidth]{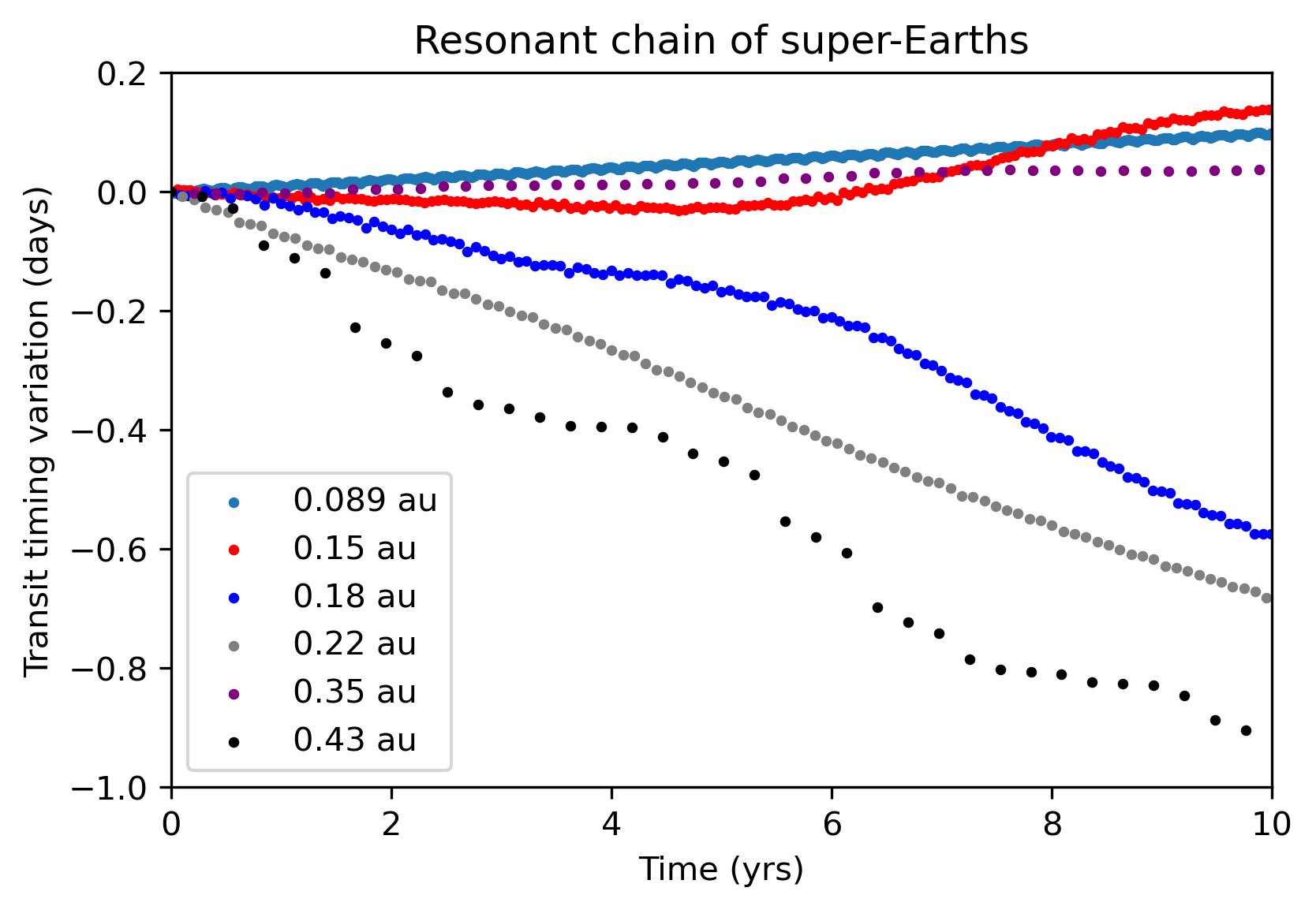} 
\caption{Transit timing variations of two systems: a 6-planet horseshoe constellation system with orbital radius of 0.25 au (left panel) and a multi-resonant super-Earth system from \citet{izidoro21}. The TTVs of all of the planets in the resonant chain were arbitrarily started at zero. In the resonant chain system, the planets' masses are (from the inside-out): $3.4 \mearth$, $7.5 \mearth$, $5.6 \mearth$, $9.2 \mearth$, $6.7 \mearth$, and $0.63 \mearth$. Pairs of neighboring planets are found in the following orbital resonances (from the inside-out): no resonance, 4:3, 4:3, 2:1, 4:3. As in Fig.~\ref{fig:TTV}, the y axis measures the deviation of the timing of each transit compared to when it would have occurred if each planet simply followed its average mean motion around the star. }
    \label{fig:TTV_short}
\end{figure*}
We performed a cursory exploration of the types of TTV signals exhibited by horseshoe constellation systems.  Figure~\ref{fig:TTV} shows examples of TTV signals from two horseshoe systems -- one with two planets and one with six planets. One can think of the change in transit timing as an azimuthal shift in each planet's position relative to an arbitrary reference direction.  In essence, the TTV signal is therefore a simple reflection of the dynamics of each horseshoe system.  In the 2-planet horseshoe system shown in Fig.~\ref{fig:234pl}, each planet's phase oscillates through roughly $180^\circ$, encounters the other planet and then returns back through $180^\circ$ to its starting point in ${\sim}100$ years.  The corresponding signal in Fig~\ref{fig:TTV} simply reflects this behavior.  

Systems with more complex dynamical evolution have correspondingly complex TTV signals.  The TTV signal of the 6-planet horseshoe system shown in Fig.~\ref{fig:TTV} shows how, after starting the simulation close together, the six planet initially spread out along their orbit (in a co-moving frame), then underwent a series of close approaches and changes in direction. To obtain a complete characterization requires observations that span the full horseshoe timescale.  However, unlike the two-planet case, the pattern never repeats exactly for $N_p > 2$.  Nonetheless, with a long-term survey of the system, a complete dynamical characterization of the co-orbital system would be possible. 

What would be the minimum observational criteria needed to characterize the 6-planet horseshoe system from Fig.~\ref{fig:TTV}, at least in broad strokes?  The first step would simply be to detect the planets in transit. If a single planet were observed in transit, the interpretation of its TTVs would be challenging.  On long timescales the planet would show large-amplitude TTVs as shown in Fig.~\ref{fig:TTV}, but it could also show wobbles on shorter timescales that could masquerade as another signal, as shown by \cite{vok14}.  If more than one of the six planets transited, it would be a challenge to determine which transit corresponded to which planet, given their equal sizes.  Assuming that barrier to be overcome, the TTV trends of the planets could be characterized after a modest number (${\sim}10$) of transit data points. These trends would be suggestive, and along with the phase information could provide a starting point for dynamical characterization.  The smoking gun for a complex horseshoe system would be an identification of the turnaround points, where a close approach between two planets reversed the sign of their relative azimuthal drift and therefore of the slope of their TTV drift.  We count 78 turnaround points over the 500 year interval shown in the 6-planet system from Fig.~\ref{fig:TTV}.  Assuming a 20 orbit window for observations at a random point within that 500 orbit interval, the odds favor the characterization of at least one of those turnaround points.  

The TTV signal of the 6-planet horseshoe system from Fig.~\ref{fig:TTV} is far more dynamic than the 2-planet horseshoe.  Given the same observing window, the odds of seeing a TTV turnaround are far higher for the 6-planet horseshoe than for the 2-planet horseshoe.  However, the much more complex dynamics of the 6-planet system would make it much harder to have confidence that the system was adequately characterized.

The duration of the TTV signal in Fig.~\ref{fig:TTV} is so long that any attempt at characterization may feel discouraging.  However, it's worth keeping in mind that the  horseshoe libration timescale scales with the orbital period. Figure~\ref{fig:TTV_short} shows the transit timing variations over a decade in two different systems on shorter-period orbits: a 6-planet horseshoe constellation with an orbital radius of 0.25 au, and a multi-resonant chain of six super-Earths extending from 0.089 au to 0.43 au that was formed in a simulation by \cite{izidoro21} that included gas-driven migration and pebble accretion.  Transit timing variations in the 6-planet horseshoe constellation are similar to those seen in Fig.~\ref{fig:TTV} but with far more frequent `turnarounds' caused by horseshoe encounters, roughly one every 1-2 years.  A decade-long survey of the system could plausibly result in a relatively complete dynamical characterization.  

In non-co-orbital systems, transit timing variations come from secular and resonant gravitational perturbations between the planets~\citep{agol05,holman05,agol17}.  TTV signals for close-in planets are generally lower-amplitude than for the co-orbital systems and are measured in minutes or hours~\citep[see][for the latest analysis of TTVs in the TRAPPIST-1 system, allowing a measurement of the planets' masses to a precision of 3-6\%]{agol21}.  The TTV signals for the different planets in the resonant chain in Fig.~\ref{fig:TTV_short} show a variety of patterns, with long-term drifts, short-term oscillations, and `chopping' patterns~\citep[visible most clearly in the planet at 0.43 au; for a description of this phenomenon, see][]{deck15}.  In contrast, the TTV signal from the horseshoe constellation shows only smooth drifts caused by relative drift in the planets' mean longitudes.  This is a key difference that could in principle allow the two types of systems to be differentiated observationally.

\section{Discussion}

\subsection{Formation of horseshoe constellations}

Co-orbital systems are a natural outcome of planet formation. The mechanism for producing co-orbitals involves gravitational scattering between neighboring protoplanets in an environment with strong damping of random velocities.  \cite{collins09} showed that co-orbital (Trojan) configurations often occur during the oligarchic phase of planetary embryo formation.  In addition, inward migration of cohorts of $\sim$Earth-mass planets frequently produces Trojan systems with similar-mass planets~\citep[in perhaps 10\% of simulations; e.g.][]{cresswell09,izidoro17}. Migration has also been shown to produce satellites on horseshoe orbits~\citep{rodriguez19}.

What conditions would be needed for the formation of a horseshoe constellation system, with many planets sharing the same orbit?  We discovered, quite by chance, that horseshoe constellations can be created by simply placing equal-mass planets along the same orbit, as long as they are not too close.  We can think of two ways in which such a situation could in principle arise naturally.  First, scattering of a large population of protoplanets could place more than two on nearby, crossing orbits. With the right amount of damping, the planets could end up in a co-orbital configuration~\citep[similar to the mechanism from ][]{collins09}.  However, if damping is too strong the planets would likely end up on Trojan (tadpole) orbits rather than horseshoes, so this mechanism would certainly be rare, as it requires a just-so scattering event followed by just the right amount of damping.  A second potential mechanism for the formation of horseshoe constellations starts from a very massive ring of dust or planetesimals.  Instablity-driven fragmentation along the ring could in principle create gravitationally-bound clumps sharing the same orbit around a star.  If those clumps produced planets of roughly equal-mass, that might lead to a horseshoe constellation system.

\subsection{Horseshoe constellations as SETI beacons}

An alternate -- and admittedly speculative -- formation scenario for horseshoe constellation systems involves their construction by a highly-advanced civilization capable of designing custom planetary systems.  In a previous paper~\citep{clement22} we argued that advanced civilizations could create complex planetary systems to add terrestrial planets in the habitable zone, or to act as beacons to mark their presence on a Galactic scale. A beacon system should appear non-natural, an exceedingly rare outcome of planet formation. In \cite{clement22} we showed how integer sequences are encoded within multi-resonant planetary systems, and that resonant chains following certain sequences (e.g., consecutive prime numbers and the Fibonacci sequence) are both unlikely to form naturally and can remain dynamically stable for long timescales.

Constellations of co-orbital planets may represent another viable orbital architecture for a SETI beacon.  We have demonstrated that horseshoe constellations are stable for long timescales for up to 24 planets per orbit, and can even persist after their host stars undergo post-main sequence evolution.  As discussed above, they also appear unlikely to represent a common outcome of planet formation.  Horseshoe constellations could also represent an intermediate step in the construction of a Dyson sphere or a Dyson swarm, as they increase the surface area along a given orbit. 

\subsection{Limitations and future work}

A main limitation of our study is that we neglected the effect of tides on the dynamics of horseshoe constellations. We could simply argue that stellar tides have at most a small effect on planets at 1 au around Sun-like stars.  However, observational biases very strongly favor the detection of planets on short-period orbits~\citep[e.g.][]{winn18}, in the orbital realm where tides are strong. Several studies have shown that tidal dissipation can affect the dynamics and stability of co-orbital systems~\citep{rodriguez13,couturier21,dobrovolskis22}. In addition, M dwarf stars are far more common in the local solar neighborhood, and would therefore provide a statistically-larger sample size and increase the chance of transit detection relative to Sun-like stars. In the future we may determine the stability of horseshoe constellations with a realistic tidal dissipation formalism, perhaps with a modified N-body code such as Mercury-T~\citep{bolmont15}. This approach would also allow us to study the dynamics of horseshoe constellations in the habitable zones of low-mass stars.

One might wonder whether horseshoe constellation systems -- or, more generally, rings of co-orbital planets with any orbital architecture -- would remain stable to perturbations, perhaps from rogue asteroids or planetary embryos.  That question is the subject of a companion study (Raymond et al, submitted).  

\section*{Acknowledgements}
We are grateful to Charles Choi, whose discussion with S.N.R. provided the seed idea for this paper and its precursor blog post.\footnote{\url{https://planetplanet.net/2020/11/19/cohorts/.}} We thank the anonymous referee for a constructive report that improved the paper. S.N.R. thanks the CNRS's PNP and MITI programs for support.  V.S.M. and S.N.R. also acknowledge support from NASA’s NExSS Virtual Planetary Laboratory, funded under NASA Astrobiology Program grant 80NSSC18K0829. A.I. acknowledges support from NASA grant 80NSSC18K0828 and the Welch Foundation grant No. C-2035-20200401.


\section*{Data Availability}

All simulations and analysis code in this paper will be made available upon reasonable request.







\appendix
\section{Integrator test}

\begin{figure}
	\includegraphics[width=\columnwidth]{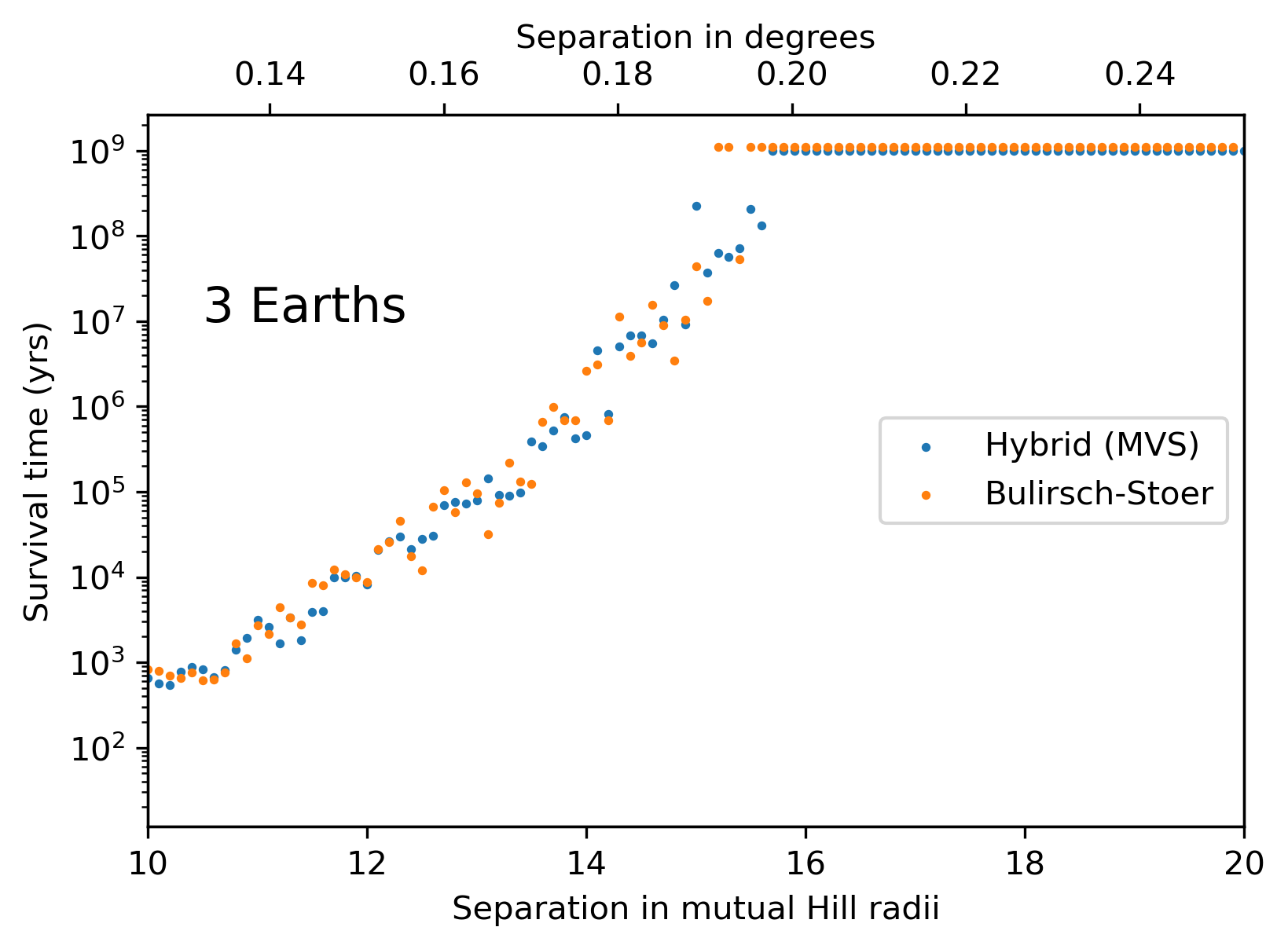}
    \caption{Stability limits for 3-Earth systems simulated with two different integration schemes: the {\tt Mercury} hybrid integrator and its Bulirsch-Stoer integrator~\citep{chambers99}. }
    \label{fig:bstest}
\end{figure}

Figure~\ref{fig:bstest} shows clear agreement between the survival of a 3-planet horseshoe system integrated with two different integration schemes: the hybrid and 'BS2' Bulirsch-stoer integrators from the {\tt Mercury} integration package~\citep{chambers99}.  

\bsp	
\label{lastpage}
\end{document}